\documentclass[aps,prd,twocolumn,superscriptaddress]{revtex4}
\usepackage{epsfig,epsf}
\usepackage{amsmath}
\usepackage{amsthm}
\usepackage{amsfonts}
\usepackage{amssymb}
\usepackage{dsfont}
\usepackage{multirow}
\usepackage{appendix}
\usepackage{slashed}
\usepackage[active]{srcltx}
\usepackage{psfrag}

\begin{document}

\title{Pseudoscalar and vector tetraquarks $bb\overline{c}\overline{c}$}
\date{\today}
\author{S.~S.~Agaev}
\affiliation{Institute for Physical Problems, Baku State University, Az--1148 Baku,
Azerbaijan}
\author{K.~Azizi}
\affiliation{Department of Physics, University of Tehran, North Karegar Avenue, Tehran
14395-547, Iran}
\affiliation{Department of Physics, Do\v{g}u\c{s} University, Dudullu-\"{U}mraniye, 34775
Istanbul, T\"{u}rkiye}
\author{H.~Sundu}
\affiliation{Department of Physics Engineering, Istanbul Medeniyet University, 34700
Istanbul, T\"{u}rkiye}

\begin{abstract}
The pseudoscalar and vector four-quark states $bb\overline{c}\overline{c}$
are studied in the context of the QCD sum rule method. We model $T_{\mathrm{%
\ \ PS}} $ and $T_{\mathrm{V}}$ as structures built of diquarks $
b^{T}C\gamma_{5}b$, $\overline{c}C\overline{c}^{T}$ and $b^{T}C\gamma _{5}b$
, $\overline{c}C\gamma_{\mu}\gamma_{5}\overline{c}^{T}$, respectively, with $
C$ being the charge conjugation matrix. The spectroscopic parameters of the
tetraquarks $T_{\mathrm{PS}}$ and $T_{\mathrm{V}}$, i.e., their masses and
current couplings are calculated using QCD two-point sum rule method. We
evaluate the full widths of $T_{\mathrm{PS}}$ and $T_{\mathrm{V}}$ by taking
into account their kinematically allowed decay channels. In the case of the
pseudoscalar particle they are processes $T_{\mathrm{PS}} \to
B_{c}^{-}B_{c}^{\ast -}$, $B_{c}^{-}B_{c}^{-}(1^{3}P_{0})$ and $B_{c}^{\ast
-}B_{c}^{-}(1^{1}P_{1})$. The vector state $T_{\mathrm{V}}$ can dissociate
to meson pairs $2 B_{c}^{-}$, $2 B_{c}^{\ast -}$ and $
B_{c}^{-}B_{c}^{-}(1^{1}P_{1})$. Partial widths of these decays are
determined by the strong couplings at relevant tetraquark-meson-meson
vertices, which evaluated in the context of the three-point sum rule
approach. Predictions obtained for the mass and full width of the
pseudoscalar $m =(13.092\pm 0.095)~\mathrm{GeV}$, $\Gamma _{\mathrm{PS}
}=(63.7\pm 13.0)~\mathrm{MeV}$ and vector $\widetilde{m} =(13.15\pm 0.10)~
\mathrm{GeV}$, $\Gamma_{\mathrm{V}}=(53.5\pm 10.3)~\mathrm{MeV}$ tetraquarks
can be useful for analyses of different four-quark resonances.
\end{abstract}

\maketitle


\section{Introduction}

\label{sec:Intro}

During past two decades four-quark mesons, i.e., tetraquarks became an
inseparable part of hadron spectroscopy. Despite the fact that already at
1970-80 years existence of exotic light mesons was predicted by researches
\cite{Jaffe:1976ig,Jaffe:1976yi}, only discoveries of charmoniumlike
resonances in last twenty years constituted this rapidly growing branch of
high energy physics. The fruitful experimental activity started from
observation of the resonance $X(3872)$ by Belle \cite{Choi:2003ue} and other
experimental collaborations \cite{Abazov:2004kp,Acosta:2003zx,Aubert:2004ns}
continues till our days. Such remarkable experimental achievements were
accompanied by theoretical progress in elaborating new and refining existing
methods and models to understand internal organization of multiquark hadrons
and explain results of measurements.

The resonances built of four heavy quarks (charm $c$ or bottom $b$ quarks)
are relatively new particles in a family of exotic mesons. They were
discovered by various experiments such as LHCb, ATLAS, and CMS in the $%
J/\psi J/\psi $ and $J/\psi \psi ^{\prime }$ spectra \cite%
{LHCb:2020bwg,Bouhova-Thacker:2022vnt,CMS:2023owd}, and occupy the mass
range $6.2-7.3~\mathrm{GeV}$. These $X$ resonances presumably contain $c$ ($%
\overline{c}$) quarks and are scalar particles. They were studied in the
context of various approaches and modeled as diquark-antidiquark states,
hadronic molecules, or interpreted as coupled-channel effects (see, for
instance, Ref.\ \cite{Agaev:2023wua} and references therein). Production
mechanisms of fully heavy tetraquarks and cross sections of relevant
processes in the LHC experiments and Future Circular Collider were
considered as well \cite{Carvalho:2015nqf,Abreu:2023wwg}.

The four-quark mesons $bb\overline{c}\overline{c}/cc\overline{b}\overline{b}$
form interesting subclass of fully heavy tetraquarks and have some
distinguished features. Namely, they bear the electric charge $\pm 2e$ and
are members of this interesting tetraquarks' family. It is worth to note
that double-charged four quark states were predicted and explored in Refs.\
\cite{Chen:2017rhl,Agaev:2017oay,Agaev:2018vag,Agaev:2021jsz}: The first
such resonance $T_{cs0}^{a}(2900)^{++}$ was discovered recently by the LHCb
Collaboration \cite{LHCb:2022xob,LHCb:2022bkt}.

Another important characteristic of the tetraquarks $bb\overline{c}\overline{%
c}/cc\overline{b}\overline{b}$ is their stability against strong two-meson
decays provided masses of these states are below relevant thresholds. In
this aspect they qualitatively differ from $cc\overline{c}\overline{c}$ or $%
bb\overline{b}\overline{b}$ tetraquarks which are strong-interaction
unstable compounds due to annihilation of constituent quark-antiquark \cite%
{Becchi:2020mjz,Becchi:2020uvq,Agaev:2023ara}. The states $bb\overline{c}%
\overline{c}/cc\overline{b}\overline{b}$ still belong to category of
hypothetical particles, but would be discovered in ongoing and/or future
experiments.

Therefore, it is worth to study these particles in a detailed form.
Actually, such investigations were performed in many theoretical
publications. In fact, spectroscopic parameters of diquark-antidiquark
states $bb\overline{c}\overline{c}$ with spin-parities $J^{\mathrm{P}}=0^{+}$%
, $1^{+}$, and $2^{+}$ were calculated in Ref.\ \cite{Wu:2016vtq}, in which
it was demonstrated that they are unstable against strong decays, because
have masses exceeding the relevant $B_{c}B_{c}$ thresholds. Unstable nature
of these states was confirmed in Ref.\ \cite{Liu:2019zuc} as well. Here,
various all-heavy tetraquarks were studied in a potential model with the
linear and Coulomb potential, and spin-spin interactions. It was shown that $%
bb\overline{c}\overline{c}$ states are approximately $300~\mathrm{MeV}$
heavier than lowest two open-flavor mesons' thresholds. Problems of fully
heavy tetraquarks with different contents and quantum numbers were also
addressed in the context of the relativistic quark model \cite%
{Galkin:2023wox}. In this framework only a state $bb\overline{c}\overline{c}$
with $J^{\mathrm{P}}=4^{+}$ was found to be stable and $33~\mathrm{MeV}$
below the $B^{\pm }(1^{3}S_{1})B^{\pm }(1^{3}D_{3})$ threshold.

The masses of the four-quark states $bb\overline{c}\overline{c}$ were
considered using QCD moment sum rule method in Ref. \cite{Wang:2021taf}, in
which they were modeled by different interpolating currents. The authors
argued that some of $J^{\mathrm{P}}=0^{+}$, $1^{+}$ and $2^{+}$ states are
strong-interaction stable structures. The same states were analyzed in the
context of the dynamical diquark model and were also found unstable against
strong decays \cite{Mutuk:2022nkw}. If exist, such structures can be
converted to standard mesons only through radiative transitions or weak
processes considered in Ref.\ \cite{Li:2019uch}.

The four-quark exotic mesons $bb\overline{c}\overline{c}$ with spin-parities
$J^{\mathrm{P}}=0^{+}$, $1^{+}$, and $2^{+}$ were explored in QCD sum rule
framework in our articles \cite{Agaev:2023tzi,Agaev:2024pej,Agaev:2024pil}.
We modeled these particles by employed different diquark and antidiquarks.
Thus, two scalar states composed of an axial-vector diquark and antidiquark,
and a pseudoscalar diquark and antidiquark were explored in Ref. \cite%
{Agaev:2023tzi} by means of the sum rule approach. We computed the masses of
these particles $(12.715\pm 0.080)~\mathrm{GeV}$ and $(13.370\pm 0.095)~%
\mathrm{GeV}$, and found that they can easily dissociate to $B_{c}^{(\ast
)-}B_{c}^{(\ast )-}$ meson pairs. The two axial-vector tetraquarks
considered in Ref.\ \cite{Agaev:2024pej} had the structures $C\sigma _{\mu
\nu }\gamma _{5}\otimes \gamma ^{\nu }C$ and $C\gamma _{\mu }\gamma
_{5}\otimes C$, respectively. The tensor tetraquark was treated as a
diquark-antidiquark state with components $b^{T}C\gamma _{\mu }b$ and $%
\overline{c}\gamma _{\nu }C\overline{c}^{T}$ \cite{Agaev:2024pil}. It should
be emphasized that none of structures studied in Refs.\ \cite%
{Agaev:2023tzi,Agaev:2024pej,Agaev:2024pil} are stable against strong
transformations. Therefore, we evaluated their full widths and classified
them as tetraquarks with modest widths.

In present work, we are going to investigate the pseudoscalar $T_{\mathrm{PS}%
}$ and vector $T_{\mathrm{V}}$ tetraquarks $bb\overline{c}\overline{c}$ by
computing their masses, current couplings, and widths. The spectroscopic
parameters of these structures are evaluated by means of the QCD sum rule
(SR) method \cite{Shifman:1978bx,Shifman:1978by}. The SR method originally
elaborated to study conventional hadrons, can also be successfully applying
to analyze parameters of multiquark particles \cite%
{Albuquerque:2018jkn,Agaev:2020zad}.

The full widths of the states $T_{\mathrm{PS}}$ and $T_{\mathrm{V}}$ are
calculated by considering their kinematically allowed decay modes. In the
case of the pseudoscalar particle they are processes $T_{\mathrm{PS}%
}\rightarrow B_{c}^{-}B_{c}^{\ast -}$, $B_{c}^{-}B_{c}^{-}(1^{3}P_{0})$ and $%
B_{c}^{\ast -}B_{c}^{-}(1^{1}P_{1})$. The vector state $T_{\mathrm{V}}$
falls apart to meson pairs $2B_{c}^{-}$, $2B_{c}^{\ast -}$ and $%
B_{c}^{-}B_{c}^{-}(1^{1}P_{1})$. Partial widths of these decays are
determined by strong couplings at relevant tetraquark-meson-meson vertices.
We evaluate these couplings in the context of the three-point sum rule
approach.

As is seen, decays of the tetraquarks $T_{\mathrm{PS}}$ and $T_{\mathrm{V}}$
at a final state contain two $B_{c}$ mesons with different spin-parities.
But, because of the small production rate, these particles are much less
studied than compared to quarkonia. The ground-level meson $B_{c}^{+}$ was
first observed by the CDF experiment at the Tevatron collider \cite%
{Abe:1998a,Abe:1998b}. A few years ago, LHC collaborations reported about
discovery of the excited states $B_{c}^{+}(2^{1}S_{0})$ and $%
B_{c}^{+}(2^{3}S_{1})$ in the invariant mass spectrum of the $B_{c}^{+}\pi
^{+}\pi ^{-}$ system \cite{ATLAS:2014lga,LHCb:2017rqe,CMS:2019uhm}. It seems
that parameters of the $B_{c}$ mesons with other quantum numbers are also
accessible and will be measured in running experiments.

This article consists of five sections. In Sec.\ \ref{sec:AV} we find the
spectroscopic parameters $m$, $\Lambda $ and $\widetilde{m}$, $\widetilde{%
\Lambda }$ of the tetraquarks $T_{\mathrm{PS}}$ and $T_{\mathrm{V}}$,
respectively. In Sec. \ref{sec:PSWidths} we evaluate full width of the
pseudoscalar state $T_{\mathrm{PS}}$ by computing partial widths of the
aforementioned processes. The $T_{\mathrm{V}}$ tetraquark's decay modes are
explored in the Sec. \ref{sec:VWidths}. We sum up and make our brief
conclusions in the last section.


\section{Spectroscopic parameters of $T_{\mathrm{PS}}$ and $T_{\mathrm{V}}$}

\label{sec:AV} 

The sum rules for the masses and current couplings (pole residues) of the
tetraquarks $T_{\mathrm{PS}}$ and $T_{\mathrm{V}}$ in the context of the QCD
sum rule method can be obtained by studying the two-point correlation
functions
\begin{equation}
\Pi (p)=i\int d^{4}xe^{ipx}\langle 0|\mathcal{T}\{J(x)J^{\dag
}(0)\}|0\rangle ,  \label{eq:CF1}
\end{equation}%
and
\begin{equation}
\Pi _{\mu \nu }(p)=i\int d^{4}xe^{ipx}\langle 0|\mathcal{T}\{J_{\mu
}(x)J_{\nu }^{\dag }(0)\}|0\rangle ,  \label{eq:CF2}
\end{equation}%
where $J(x)$ and $J_{\mu }(x)$ are interpolating currents for the
pseudoscalar and vector tetraquarks, respectively.

In general, $T_{\mathrm{PS}}$ and $T_{\mathrm{V}}$ can be modeled by means
of different diquark and antidiquark components \cite{Jaffe:2004ph,Du:2012wp}%
. One may construct relevant interpolating currents using diquarks both
symmetric and antisymmetric in color indices. In the case of the $\mathrm{PS}
$ tetraquarks $C\gamma _{5}\otimes C$ and $C\otimes \gamma _{5}C$ are color
symmetric structures, whereas $C\sigma _{\mu \nu }\otimes \sigma ^{\mu \nu
}\gamma _{5}C$ has the color antisymmetric organization \cite{Wang:2021taf}.
Among the four possible currents for the vector tetraquark $T_{\mathrm{V}}$,
structures, for example, $C\gamma _{5}\otimes \gamma _{\mu }\gamma _{5}C$
and $C\sigma _{\mu \nu }\otimes \gamma ^{\nu }C$ are symmetric and
antisymmetric in color indices, respectively.

We consider states composed of a scalar diquark and pseudoscalar antidiquark
(for $T_{\mathrm{PS}}$ ) and scalar diquark and vector antidiquark (for $T_{%
\mathrm{V}}$). In accordance with our models, the interpolating currents for
$T_{\mathrm{PS}}$ and $T_{\mathrm{V}}$ are represented by the expressions:
\begin{equation}
J(x)=[b_{a}^{T}(x)C\gamma _{5}b_{b}(x)][\overline{c}_{a}(x)C\overline{c}%
_{b}^{T}(x)],  \label{eq:C1}
\end{equation}%
for the tetraquark $T_{\mathrm{PS}}$ and
\begin{equation}
J_{\mu }(x)=[b_{a}^{T}(x)C\gamma _{5}b_{b}(x)][\overline{c}_{a}(x)\gamma
_{\mu }\gamma _{5}C\overline{c}_{b}^{T}(x)],  \label{eq:C2}
\end{equation}%
for $T_{\mathrm{V}}$. These currents belong to the symmetric $\mathbf{[6]}%
_{bb}\otimes \mathbf{[}\overline{\mathbf{6}}\mathbf{]}_{\overline{c}%
\overline{c}}$ representations of the color group $SU_{c}(3)$. They are
written down in the compact form bearing in mind that second terms in color
symmetric diquark and antidiquark fields, for instance, in $\overline{c}%
_{a}(x)C\overline{c}_{b}^{T}(x)+\overline{c}_{b}(x)C\overline{c}_{a}^{T}(x)$
[see, Eq.\ (\ref{eq:C1})] lead to the same contributions as the first ones.
In the framework of SR method, $J(x)$ and $J_{\mu }(x)$ describe tetraquarks
with lowest masses among pseudoscalar and vector exotic $bb\overline{c}%
\overline{c}$ mesons, whereas the states with structures $[\overline{\mathbf{%
3}}]_{bb}\otimes \lbrack \mathbf{3}]_{\overline{c}\overline{c}}$ have higher
masses \cite{Wang:2021taf}.

In what follows, we are going to present in an extended format calculations
of the $T_{\mathrm{PS}}$ exotic meson's parameters, and write down principal
formulas in the case of the tetraquark $T_{\mathrm{V}}.$To derive the sum
rules for the mass $m$ and current coupling $\Lambda $ of the pseudoscalar
tetraquark, we first write down the correlation function $\Pi (p)$ using
physical parameters of the particle $T_{\mathrm{PS}}$. To this end, it is
necessary to insert into Eq.\ (\ref{eq:CF1}) a full set of intermediate
states with the content and spin-parity of the tetraquark $T_{\mathrm{PS}}$,
and carry out integration over $x$. After these operations, we get
\begin{equation}
\Pi ^{\mathrm{Phys}}(p)=\frac{\langle 0|J|T_{\mathrm{PS}}(p)\rangle \langle
T_{\mathrm{PS}}(p)|J^{\dagger }|0\rangle }{m^{2}-p^{2}}+\cdots .
\label{eq:CF1a}
\end{equation}%
Here, the term written down explicitly is a contribution of the ground-state
particle. By the dots, we denote effects due to higher resonances and
continuum states.

Having introduced the matrix element
\begin{equation}
\langle 0|J|T_{\mathrm{PS}}(p)\rangle =\Lambda  \label{eq:MElement1}
\end{equation}%
one can considerably simplify the correlator $\Pi ^{\mathrm{Phys}}(p)$, and
find%
\begin{equation}
\Pi ^{\mathrm{Phys}}(p)=\frac{|\Lambda |^{2}}{m^{2}-p^{2}}+\cdots .
\label{eq:PhysSide}
\end{equation}%
The function $\Pi ^{\mathrm{Phys}}(p)$ contains only the Lorentz structure $%
\sim \mathrm{I}$, therefore $|\Lambda |^{2}/(m^{2}-p^{2})$ is the invariant
amplitude $\Pi ^{\mathrm{Phys}}(p^{2})$ which will be employed in the
following analysis.

The SR studies require computations of the correlation function with some
fixed accuracy by applying the operator product expansion ($\mathrm{OPE}$).
Our computations are performed by taking into account dimension-$4$ gluon
condensate $\langle \alpha _{s}G^{2}/\pi \rangle $. To find $\Pi ^{\mathrm{%
OPE}}(p)$, we insert the current $J(x)$ into Eq.\ (\ref{eq:CF1}), contract
relevant quark fields and express them in terms of the heavy quark
propagators. After these manipulations the correlator $\Pi ^{\mathrm{OPE}%
}(p) $ reads
\begin{eqnarray}
&&\Pi ^{\mathrm{OPE}}(p)=i\int d^{4}xe^{ipx}\left\{ \left[ \mathrm{Tr}\left(
\gamma _{5}\widetilde{S}_{b}^{aa^{\prime }}(x)\gamma _{5}S_{b}^{bb^{\prime
}}(x)\right) \right. \right.  \notag \\
&&\left. +\mathrm{Tr}\left( \gamma _{5}\widetilde{S}_{b}^{ba^{\prime
}}(x)\gamma _{5}S_{b}^{ab^{\prime }}(x)\right) \right] \left[ \mathrm{Tr}%
\left( S_{c}^{a^{\prime }a}(-x)\widetilde{S}_{c}^{b^{\prime }b}(-x)\right)
\right.  \notag \\
&&\left. \left. +\mathrm{Tr}\left( S_{c}^{b^{\prime }a}(-x)\widetilde{S}%
_{c}^{a^{\prime }b}(-x)\right) \right] \right\} ,  \label{eq:QCDSide1}
\end{eqnarray}%
where
\begin{equation}
\widetilde{S}_{b(c)}(x)=CS_{b(c)}(x)C.
\end{equation}%
Here, $S_{b}(x)$ and $S_{c}(x)$ are the propagators of $b$ and $c$ quarks
explicit expressions of which are presented in Appendix. The correlator $\Pi
^{\mathrm{OPE}}(p)$ has also simple structure proportional to $\mathrm{I.}$
We denote the corresponding invariant amplitude by $\Pi ^{\mathrm{OPE}%
}(p^{2})$ and employ it to derive SRs for $m$ and $\Lambda $.

\bigskip\ The function $\Pi ^{\mathrm{Phys}}(p^{2})$ can be expressed as the
dispersion integral%
\begin{equation}
\Pi ^{\mathrm{Phys}}(p^{2})=\int_{4\mathcal{M}^{2}}^{\infty }\frac{\rho ^{%
\mathrm{Phys}}(s)ds}{s-p^{2}}+\cdots ,  \label{eq:DisRel}
\end{equation}%
where $\mathcal{M=}m_{b}+m_{c}$, and dots indicate subtraction terms
necessary to make the whole expression finite. The spectral density $\rho ^{%
\mathrm{Phys}}(s)$ is equal to the imaginary part of $\Pi ^{\mathrm{Phys}%
}(p) $,%
\begin{equation}
\rho ^{\mathrm{Phys}}(s)=\frac{1}{\pi }\mathrm{Im}\Pi ^{\mathrm{Phys}%
}(s)=|\Lambda |^{2}\delta (s-m^{2})+\rho ^{\mathrm{h}}(s)\theta(s-s_0),
\label{eq:SDensity}
\end{equation}%
where $ s_0 $ is the continuum subtraction parameter.
Here, contribution of the ground-level particle, i.e., the pole term is
separated from effects of higher resonances and continuum states which are
characterized by an unknown hadronic spectral density $\rho ^{\mathrm{h}}(s)$%
. It is easy to see that $\rho ^{\mathrm{Phys}}(s)$ substituted into Eq.\ (%
\ref{eq:DisRel}) leads to the expression 
\begin{equation}
\Pi ^{\mathrm{Phys}}(p^{2})=\frac{|\Lambda |^{2}}{m^{2}-p^{2}}+\int_{s_0}^{\infty }\frac{\rho ^{\mathrm{h}}(s)ds}{s-p^{2}}.
\label{eq:InvAmp2}
\end{equation}

The amplitude $\Pi ^{\mathrm{OPE}}(p^{2})$ can be computed theoretically in
deep Euclidean region $p^{2}\ll 0$ using the operator product expansion. The
coefficient functions in $\mathrm{OPE}$ could be obtained using methods of
perturbative QCD, whereas nonperturbative information is contained in vacuum
expectation values of various quark, gluon and mixed operators. In the case
under consideration $\Pi ^{\mathrm{OPE}}(p^{2})$ contains only the gluon
matrix element $\langle \alpha _{s}G^{2}/\pi \rangle $.

Having continued $\Pi ^{\mathrm{OPE}}(p^{2})$ analytically to the Minkowski
domain and calculated its imaginary part, we find the two-point spectral
density $\rho ^{\mathrm{OPE}}(s)$. In the region $p^{2}\ll 0$ we apply the
Borel transformation to remove subtraction terms in the dispersion integral
and suppress contributions of higher resonances and continuum states. In the
case of $\Pi ^{\mathrm{Phys}}(p^{2})$, we find%
\begin{equation}
\mathcal{B}\Pi ^{\mathrm{Phys}}(p^{2})=|\Lambda |^{2}e^{-m^{2}/M^{2}}+\int_{s_0}^{\infty }ds\rho ^{\mathrm{h}}(s)e^{-s/M^{2}},
\label{eq:CorBor}
\end{equation}%
where $M^{2}$ is the Borel parameter. The dispersion representation in terms
of $\rho ^{\mathrm{OPE}}(s)$ can be written down for $\Pi ^{\mathrm{OPE}%
}(p^{2})$ as well. Afterwards, by equating the Borel transformations for $%
\Pi ^{\mathrm{Phys}}(p^{2})$ and $\Pi ^{\mathrm{OPE}}(p^{2})$, employing the
assumption about hadron-parton duality and matching $\rho ^{\mathrm{h}%
}(s)\simeq \rho ^{\mathrm{OPE}}(s)$ in duality region, we subtract second
term in Eq.\ (\ref{eq:CorBor}) from the QCD side of the obtained equality
and get
\begin{equation}
|\Lambda |^{2}e^{-m^{2}/M^{2}}=\Pi (M^{2},s_{0}),  \label{eq:SR}
\end{equation}%
where
\begin{equation}
\Pi (M^{2},s_{0})=\int_{4\mathcal{M}^{2}}^{s_{0}}ds\rho ^{\mathrm{OPE}%
}(s)e^{-s/M^{2}}+\Pi (M^{2}). \label{eq:CorrF}
\end{equation}%
 The second component $\Pi
(M^{2})$ in Eq.\ (\ref{eq:CorrF}) contains nonperturbative contributions
computed directly from $\Pi ^{\mathrm{OPE}}(p)$. Explicit expression for $%
\Pi (M^{2},s_{0})$ is presented in Appendix.

We see that the parameters $m$ and $\Lambda $ of the tetraquark $T_{\mathrm{%
PS}}$ are written down in terms of $\rho ^{\mathrm{OPE}}(s)$ and $\Pi (M^{2})
$ which are calculated in the quark-gluon framework. The second equation
which is required to find SRs for the mass and coupling of the tetraquark
can be derived by applying the operator $d/d(-1/M^{2})$ to both sides of the
expression Eq.\ (\ref{eq:SR}). By solving the system of equations obtained
by this way, we get

\begin{equation}
m^{2}=\frac{\Pi ^{\prime }(M^{2},s_{0})}{\Pi (M^{2},s_{0})},
\label{eq:SRMass}
\end{equation}%
and
\begin{equation}
|\Lambda |^{2}=e^{m^{2}/M^{2}}\Pi (M^{2},s_{0}),  \label{eq:SRCoupl}
\end{equation}%
where $\Pi ^{\prime }(M^{2},s_{0})=d\Pi (M^{2},s_{0})/d(-1/M^{2})$.

To perform numerical computations, we use the following input parameters
\begin{eqnarray}
&&\langle \alpha _{s}G^{2}/\pi \rangle =(0.012\pm 0.004)~\mathrm{GeV}^{4},
\notag \\
&&\overline{m}_{b}(\mu =\overline{m}_{b})=4.18_{-0.02}^{+0.03}~\mathrm{GeV},
\notag \\
&&\overline{m}_{c}(\mu =\overline{m}_{c})=(1.27\pm 0.02)~\mathrm{GeV}.
\label{eq:Parametr}
\end{eqnarray}%
Here, $\overline{m}_{b}$ and $\overline{m}_{c}$ correspond to the running $b$
and $c$ quark masses in the $\overline{\mathrm{MS}}$ scheme at the scales $%
\mu =\overline{m}_{b}$ and $\mu =\overline{m}_{c}$ \cite{PDG:2022},
respectively. The gluon condensate $\langle \alpha _{s}G^{2}/\pi \rangle $
was extracted from analysis of hadronic processes \cite%
{Shifman:1978bx,Shifman:1978by}.

Numerical analysis requires fulfillment of some necessary conditions.
Because $M^{2}$ and $s_{0}$ are auxiliary parameters, they have to comply
with constraints of the SR analysis. The dominance of the pole contribution (%
$\mathrm{PC}$)
\begin{equation}
\mathrm{PC}=\frac{\Pi (M^{2},s_{0})}{\Pi (M^{2},\infty )},  \label{eq:PC}
\end{equation}%
is among important requirements of our studies. Another restriction imposed
on $M^{2}$ and $s_{0}$ is connected with convergence of the operator product
expansion. In the case of fully heavy tetraquarks the correlator $\Pi
(M^{2},s_{0})$ depends only on gluon condensates, and does not contain light
quark vacuum expectation values. In the present work we take into account,
due to smallness of gluon condensates, a contribution of the dimension-$4$
term $\sim \langle \alpha _{s}G^{2}/\pi \rangle $, therefore it is enough to
demand fulfilment of the restriction $|\Pi ^{\mathrm{Dim4}%
}(M^{2},s_{0})|\leq 0.05\cdot \Pi (M^{2},s_{0})$. The stability of physical
quantities extracted from the sum rules upon variation of the Borel
parameter $M^{2}$ should be met as well.

\begin{figure}[h]
\includegraphics[width=8.5cm]{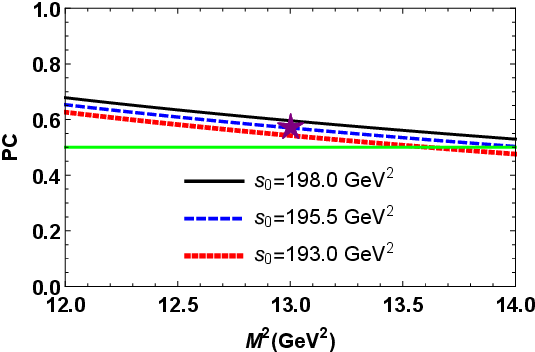}
\caption{The pole contribution $\mathrm{PC}$ as a function of $M^{2}$ at
fixed $s_{0}$. The red star marks the point $M^{2}=13~\mathrm{GeV}^{2}$ and $%
s_{0}=195.5~\mathrm{GeV}^{2}$. }
\label{fig:PC}
\end{figure}

Our calculations demonstrate that regions
\begin{equation}
M^{2}\in \lbrack 12,14]~\mathrm{GeV}^{2},\ s_{0}\in \lbrack 193,198]~\mathrm{%
GeV}^{2},  \label{eq:Wind1}
\end{equation}%
satisfy all necessary constraints imposed on the parameters $M^{2}$ and $%
s_{0}$. In fact, these working windows ensure the dominance of $\mathrm{PC}$%
, which is seen in Fig.\ \ref{fig:PC}: At $M^{2}=12~\mathrm{GeV}^{2}$ and $%
14~\mathrm{GeV}^{2}$ the pole contribution is equal to $0.65$ and $0.5$,
respectively. The nonperturbative dimension-$4$ contribution at $M^{2}=12~%
\mathrm{GeV}^{2}$ constitutes $-1.8\%$ of the function $\Pi (M^{2},s_{0})$
in agreement with the constraint introduced above.

The SR requirements depend also on the continuum subtraction parameter $%
s_{0} $. The working windows in Eq.\ (\ref{eq:Wind1}) are obtained by
simultaneous variation of the parameters $M^{2}$ and $s_{0}$. Additionally,
self-consistency of performed analysis implies fulfillment of the condition $%
m<\sqrt{s_{0}}$, i.e., the mass of the particle extracted from SR should not
exceed $\sqrt{s_{0}}$. The region for $s_{0}$ bears also information about
the mass of a first radially excited tetraquark $T_{\mathrm{PS}}(2S)$. In
fact, because we investigate the ground-level tetraquark $T_{\mathrm{PS}}$,
the parameter $s_{0}$ separates it from higher resonances and continuum
states. The excited state $T_{\mathrm{PS}}(2S)$ is the first particle in
category of "higher resonances" and its mass $m^{\ast }$ should obey a
restriction $m^{\ast }\geq $ $\sqrt{s_{0}}$. In the case of conventional
hadrons, as usual, we know masses of excited particles and can easily
control $s_{0}$. In the case of tetraquarks, the parameter $s_{0}$ provides
lower limit for $m^{\ast }$ allowing one to estimate it.

The SRs for the mass $m$ and current coupling $|\Lambda |$ give%
\begin{eqnarray}
m &=&(13.092\pm 0.095)~\mathrm{GeV},  \notag \\
|\Lambda | &=&(2.26\pm 0.26)~\mathrm{GeV}^{5}.  \label{eq:Results1}
\end{eqnarray}%
Let us note that $m$ and $|\Lambda |$ in Eq.\ (\ref{eq:Results1}) are mean
values of these parameters averaged over the regions in Eq.\ (\ref{eq:Wind1}%
). They effectively correspond to SR predictions at $M^{2}=13~\mathrm{GeV}%
^{2}$ and $s_{0}=195.5~\mathrm{GeV}^{2}$ where $\mathrm{PC}=0.57$. The mass $%
m$ is smaller than $\sqrt{s_{0}}\approx 13.98~\mathrm{GeV}$ and $m^{\ast
}\geq 13.98~\mathrm{GeV}$, as it has been just discussed above. The mass gap
$m^{\ast }-m\approx $ $900~\mathrm{MeV}$ is a reasonable estimate if one
takes into account $m[B_{c}(2S)]-m[B_{c}(1S)]\approx $ $600~\mathrm{MeV}$
and quark content of these ordinary mesons. The mass $m$ of the tetraquark $%
T_{\mathrm{PS}}$ as a function of $M^{2}$ and $s_{0}$ is plotted in Fig.\ %
\ref{fig:Mass1}.

Investigation of the vector tetraquark $T_{\mathrm{V}}$ differs from
analysis of $T_{\mathrm{PS}}$ by nonessential technical details. Indeed, the
correlator $\Pi _{\mu \nu }^{\mathrm{Phys}}(p)$ for the vector tetraquark $%
T_{\mathrm{V}}$ is determined by the expression
\begin{equation}
\Pi _{\mu \nu }^{\mathrm{Phys}}(p)=\frac{|\widetilde{\Lambda }|^{2}}{%
\widetilde{m}^{2}-p^{2}}\left( -g_{\mu \nu }+\frac{p_{\mu }p_{\nu }}{%
\widetilde{m}^{2}}\right) +\cdots ,  \label{eq:PhysSide1}
\end{equation}%
where $\widetilde{m}$ and $\widetilde{\Lambda }$ are the mass and current
coupling of $T_{\mathrm{V}}$, respectively. To derive Eq.\ (\ref%
{eq:PhysSide1}), we have used the matrix element
\begin{equation}
\langle 0|J_{\mu }|T_{\mathrm{V}}(p,\epsilon )\rangle =\widetilde{\Lambda }%
\epsilon _{\mu }(p),  \label{eq:ME1}
\end{equation}%
with $\epsilon _{\mu }(p)$ being the polarization vector of $T_{\mathrm{V}}$%
. We employ the invariant amplitude $\widetilde{\Pi }^{\mathrm{Phys}}(p^{2})$
which corresponds to the term $\sim g_{\mu \nu }$, because it receives
contributions only from the spin-$1$ particle.

The QCD side of the sum rules $\Pi _{\mu \nu }^{\mathrm{OPE}}(p)$ has the
following form

\begin{widetext}

\begin{figure}[h!]
\begin{center}
\includegraphics[totalheight=6cm,width=8cm]{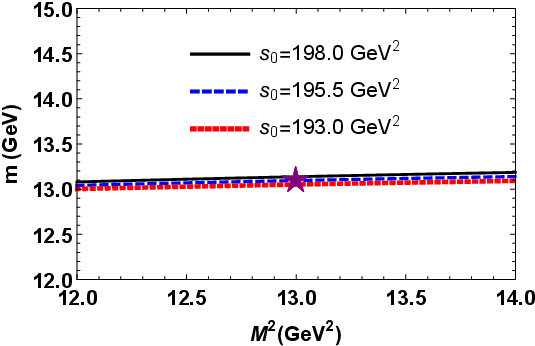}
\includegraphics[totalheight=6cm,width=8cm]{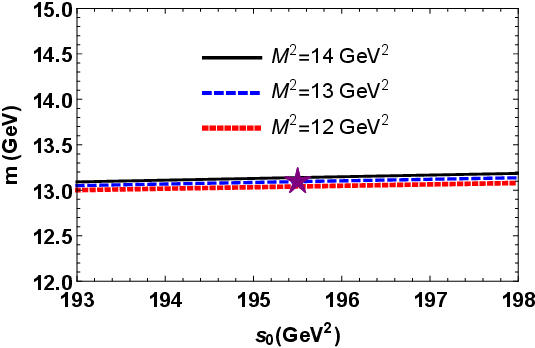}
\end{center}
\caption{Mass $m$ of the tetraquark $T_{\mathrm{PS}}$ vs $M^{2}$ and vs $s_0$ parameters (left and right panels, respectively). The stars fix the point $%
M^{2}=13~\mathrm{GeV}^{2}$ and $s_{0}=195.5~\mathrm{GeV}^{2}$.}
\label{fig:Mass1}
\end{figure}

\begin{figure}[h!]
\begin{center}
\includegraphics[totalheight=6cm,width=8cm]{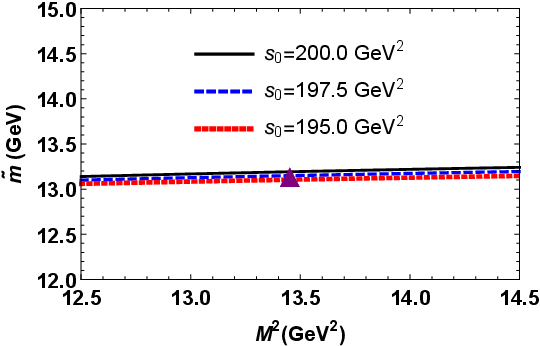}
\includegraphics[totalheight=6cm,width=8cm]{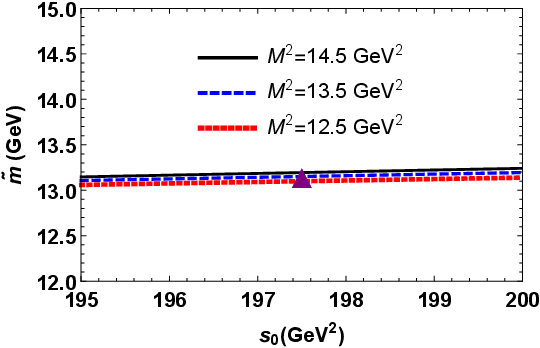}
\end{center}
\caption{Dependence of the mass $\widetilde{m}$ on the Borel parameter $M^{2}$ (left panel), and continuum threshold parameter $s_0$ (right panel). The triangles show the point $%
M^{2}=13.5~\mathrm{GeV}^{2}$ and $s_{0}=197.5~\mathrm{GeV}^{2}$, where the
mass $\widetilde{m}$ and current coupling $\widetilde{\Lambda}$ have
effectively been evaluated.}
\label{fig:Mass2}
\end{figure}

\end{widetext}

\begin{eqnarray}
&&\Pi _{\mu \nu }^{\mathrm{OPE}}(p)=i\int d^{4}xe^{ipx}\left\{ \left[
\mathrm{Tr}\left( S_{c}^{a^{\prime }a}(-x)\gamma _{\mu }\gamma _{5}%
\widetilde{S}_{c}^{b^{\prime }b}(-x)\right. \right. \right.  \notag \\
&&\left. \left. \times \gamma _{5}\gamma _{\nu }\right) +\mathrm{Tr}\left(
S_{c}^{b^{\prime }a}(-x)\gamma _{\mu }\gamma _{5}\widetilde{S}%
_{c}^{a^{\prime }b}(-x)\gamma _{5}\gamma _{\nu }\right) \right]  \notag \\
&&\times \left[ \mathrm{Tr}\left( \gamma _{5}\widetilde{S}_{b}^{aa^{\prime
}}(x)\gamma _{5}S_{b}^{bb^{\prime }}(x)\right) +\mathrm{Tr}\left( \gamma _{5}%
\widetilde{S}_{b}^{ba^{\prime }}(x)\right. \right.  \notag \\
&&\left. \left. \left. \times \gamma _{5}S_{b}^{ab^{\prime }}(x)\right)
\right] \right\} .  \label{eq:QCDSide2}
\end{eqnarray}%
To continue analysis, we choose the term $\sim g_{\mu \nu }$ in the
correlation function $\Pi _{\mu \nu }^{\mathrm{OPE}}(p)$ and denote by $%
\widetilde{\Pi }^{\mathrm{OPE}}(p^{2})$ the corresponding invariant
amplitude.

The numerical analyses are performed by employing the following working
windows for the Borel and continuum subtraction parameters
\begin{equation}
M^{2}\in \lbrack 12.5,14.5]~\mathrm{GeV}^{2},\ s_{0}\in \lbrack 195,200]~%
\mathrm{GeV}^{2}.  \label{eq:Wind1A}
\end{equation}%
These regions meet all standard restrictions of the SR computations. Thus,
the pole contribution on the average in $s_{0}$ at $M^{2}=14.5~\mathrm{GeV}%
^{2}$ and $M^{2}=12.5~\mathrm{GeV}^{2}$ is equal to $\mathrm{PC}\approx 0.5$
and $\mathrm{PC}$ $\approx 0.63$, respectively. At $M^{2}=12.5~\mathrm{GeV}%
^{2}$ the $\mathrm{Dim}4$ term is negative and equal to $2.2\%$ of the full
result.

The mass $\widetilde{m}$ and current coupling $|\widetilde{\Lambda }|$ are
equal to
\begin{eqnarray}
\widetilde{m} &=&(13.15\pm 0.10)~\mathrm{GeV},  \notag \\
|\widetilde{\Lambda }| &=&(2.31\pm 0.26)~\mathrm{GeV}^{5}.
\label{eq:Result1}
\end{eqnarray}%
The dependence of $\widetilde{m}$ on the Borel and continuum subtraction
parameters is drawn in Fig.\ \ref{fig:Mass2}.


\section{Decays of the tetraquark $T_{\mathrm{PS}}$}

\label{sec:PSWidths}


The mass $m$ of the diquark-antidiquark state $T_{\mathrm{PS}}$ fixes its
possible decay channels $T_{\mathrm{PS}}\rightarrow B_{c}^{-}B_{c}^{\ast -}$%
, $B_{c}^{-}B_{c}^{-}(1^{3}P_{0})$ and $B_{c}^{\ast -}B_{c}^{-}(1^{1}P_{1})$%
. To estimate the thresholds for production of these meson pairs, we employ
the experimental information on the mass $m_{B_{c}}=(6274.47\pm 0.27)~%
\mathrm{MeV}$ of the meson $B_{c}^{-}$ \cite{PDG:2022}. For the masses of
the vector $B_{c}^{\ast -}$, scalar $B_{c}^{-}(1^{3}P_{0})$ and axial-vector
$B_{c}^{-}(1^{1}P_{1})$ mesons, we use the theoretical predictions $%
m_{1}[B_{c}^{\ast }]=6338~\mathrm{MeV}$, $m_{2}[B_{c}(1^{3}P_{0})]=6706~%
\mathrm{MeV}$ and $m_{3}[B_{c}(1^{1}P_{1})]=6750~\mathrm{MeV}$ from Ref.\
\cite{Godfrey:2004ya}. It is clear, that decays $T_{\mathrm{PS}}\rightarrow
B_{c}^{-}B_{c}^{\ast -}$, $B_{c}^{-}B_{c}^{-}(1^{3}P_{0})$ and $B_{c}^{\ast
-}B_{c}^{-}(1^{1}P_{1})$ are kinematically allowed processes for the
tetraquark $T_{\mathrm{PS}}$.


\subsection{Process $T_{\mathrm{PS}}\rightarrow B_{c}^{-}B_{c}^{\ast -}$}


The partial width of the process $T_{\mathrm{PS}}\rightarrow
B_{c}^{-}B_{c}^{\ast -}$ besides input parameters is determined by the
strong coupling $g_{1}$ of particles at the tetraquark-meson-meson vertex $%
T_{\mathrm{PS}}B_{c}^{-}B_{c}^{\ast -}$. To evaluate $g_{1}$, it is
necessary to explore the three-point correlation function
\begin{eqnarray}
\Pi _{\mu }(p,p^{\prime }) &=&i^{2}\int d^{4}xd^{4}ye^{ip^{\prime
}y}e^{-ipx}\langle 0|\mathcal{T}\{J_{\mu }^{1}(y)  \notag \\
&&\times J^{B_{c}}(0)J^{\dag }(x)\}|0\rangle ,  \label{eq:CF3}
\end{eqnarray}%
which allow us to find the form factor $g_{1}(q^{2})$. The latter at the
mass shell $q^{2}=m_{B_{c}}^{2}$ of the meson $B_{c}^{-}$ is equal to the
coupling $g_{1}$.

Above, $J(x)$ is given by Eq.\ (\ref{eq:C1}), whereas interpolating currents
$J^{B_{c}}(x)$ and $J_{\mu }^{1}(x)$ for the mesons $B_{c}^{-}$ and $%
B_{c}^{\ast -}$ are defined by the expressions
\begin{eqnarray}
J^{B_{c}}(x) &=&\overline{c}_{j}(x)i\gamma _{5}b_{j}(x),  \notag \\
J_{\mu }^{1}(x) &=&\overline{c}_{i}(x)\gamma _{\mu }b_{i}(x).  \label{eq:CR3}
\end{eqnarray}

In order to study the correlation function $\Pi _{\mu }(p,p^{\prime })$ in
the QCD sum rule framework, we first have to express $\Pi _{\mu
}(p,p^{\prime })$ using physical parameters of the particles involved into
the decay. Then, the correlator $\Pi _{\mu }^{\mathrm{Phys}}(p,p^{\prime })$
which forms the physical side of SR, becomes equal to
\begin{eqnarray}
&&\Pi _{\mu }^{\mathrm{Phys}}(p,p^{\prime })=\frac{\langle 0|J_{\mu
}^{1}|B_{c}^{\ast }(p^{\prime },\varepsilon )\rangle }{p^{\prime 2}-m_{1}^{2}%
}\frac{\langle 0|J^{B_{c}}|B_{c}(q)\rangle }{q^{2}-m_{B_{c}}^{2}}  \notag \\
&&\times \langle B_{c}^{\ast }(p^{\prime },\varepsilon )B_{c}(q)|T_{\mathrm{%
PS}}(p)\rangle \frac{\langle T_{\mathrm{PS}}(p)|J^{\dag }|0\rangle }{%
p^{2}-m^{2}}  \notag \\
&&+\cdots .  \label{eq:CF5}
\end{eqnarray}%
Here the contribution of the ground-level states is presented explicitly:
Other contributions are shown by the ellipses.

To rewrite $\Pi _{\mu }^{\mathrm{Phys}}(p,p^{\prime })$, we introduce the
matrix elements
\begin{eqnarray}
&&\langle 0|J^{B_{c}}|B_{c}\rangle =\frac{f_{B_{c}}m_{B_{c}}^{2}}{m_{b}+m_{c}%
},  \notag \\
&&\langle 0|J_{\mu }^{1}|B_{c}^{\ast }(p^{\prime },\varepsilon )\rangle
=f_{1}m_{1}\varepsilon _{\mu }(p^{\prime }),  \label{eq:ME2}
\end{eqnarray}%
with $\varepsilon _{\mu }$ being the polarization vector of the meson $%
B_{c}^{\ast -}$. Here, $f_{B_{c}}=(371\pm 37)~\mathrm{MeV}$ and $f_{1}=471~%
\mathrm{MeV}$ are the decay constants of the mesons $B_{c}^{-}$ and $%
B_{c}^{\ast -}$ \cite{Wang:2024fwc,Eichten:2019gig}, respectively. The
vertex $T_{\mathrm{PS}}B_{c}^{-}B_{c}^{\ast -}$ is modeled as
\begin{equation}
\langle B_{c}^{\ast }(p^{\prime },\varepsilon )B_{c}(q)|T_{\mathrm{PS}%
}(p)\rangle =g_{1}(q^{2})\varepsilon ^{\ast }\cdot p.
\end{equation}%
Having introduced the necessary matrix elements, it is easy to get
\begin{eqnarray}
&&\Pi _{\mu }^{\mathrm{Phys}}(p,p^{\prime })=g_{1}(q^{2})\frac{|\Lambda
|f_{B_{c}}m_{B_{c}}^{2}f_{1}m_{1}}{(m_{b}+m_{c})\left( p^{2}-m^{2}\right)
\left( p^{\prime 2}-m_{1}^{2}\right) }  \notag \\
&&\times \frac{1}{(q^{2}-m_{B_{c}}^{2})}\left( \frac{m^{2}+m_{1}^{2}-q^{2}}{%
2m_{1}^{2}}p_{\mu }^{\prime }-p_{\mu }\right) +\cdots .  \label{eq:CF6}
\end{eqnarray}%
The correlator $\Pi _{\mu }^{\mathrm{Phys}}(p,p^{\prime })$ contains two
different Lorentz structures: In order to derive SR for $g_{1}(q^{2})$, we
prefer to work with a term proportional to $p_{\mu }^{\prime }$, and denote
the corresponding invariant amplitude by $\Pi _{1}^{\mathrm{Phys}%
}(p^{2},p^{\prime 2},q^{2})$.

The correlation function $\Pi _{\mu }^{\mathrm{OPE}}(p,p^{\prime })$,
calculated using the quark-gluon degrees of freedom, reads
\begin{eqnarray}
&&\Pi _{\mu }^{\mathrm{OPE}}(p,p^{\prime })=2i\int d^{4}xd^{4}ye^{ip^{\prime
}y}e^{-ipx}\left\{ \mathrm{Tr}\left[ \gamma _{\mu }S_{b}^{ia}(y-x)\right.
\right.  \notag \\
&&\left. \times \gamma _{5}\widetilde{S}_{b}^{jb}(-x)\gamma _{5}\widetilde{S}%
_{c}^{bj}(x)S_{c}^{ai}(x-y)\right]  \notag \\
&&\left. +\mathrm{Tr}\left[ \gamma _{\mu }S_{b}^{ia}(y-x)\gamma _{5}%
\widetilde{S}_{b}^{jb}(-x)\gamma _{5}\widetilde{S}_{c}^{aj}(x)S_{c}^{bi}(x-y)%
\right] \right\} .  \notag \\
&&  \label{eq:QCDside2}
\end{eqnarray}%
The $\Pi _{\mu }^{\mathrm{OPE}}(p,p^{\prime })$ contains two Lorentz
structures proportional to $p_{\mu }^{\prime }$ and $p_{\mu }$, as well.
Having labeled by $\Pi _{1}^{\mathrm{OPE}}(p^{2},p^{\prime 2},q^{2})$ the
amplitude that corresponds to the term $p_{\mu }^{\prime }$, we get the SR
for the form factor $g_{1}(q^{2})$
\begin{eqnarray}
&&g_{1}(q^{2})=\frac{2(m_{b}+m_{c})m_{1}}{|\Lambda
|f_{B_{c}}m_{B_{c}}^{2}f_{1}}\frac{q^{2}-m_{B_{c}}^{2}}{m^{2}+m_{1}^{2}-q^{2}%
}  \notag \\
&&\times e^{m^{2}/M_{1}^{2}}e^{m_{1}^{2}/M_{2}^{2}}\Pi _{1}(\mathbf{M}^{2},%
\mathbf{s}_{0},q^{2}).  \label{eq:SRCoup2}
\end{eqnarray}%
In Eq.\ (\ref{eq:SRCoup2}), $\Pi _{1}(\mathbf{M}^{2},\mathbf{s}_{0},q^{2})$
is the Borel transformed and subtracted function $\Pi _{1}^{\mathrm{OPE}%
}(p^{2},p^{\prime 2},q^{2})$. It depends on the quantities $\mathbf{M}%
^{2}=(M_{1}^{2},M_{2}^{2})$ and $\mathbf{s}_{0}=(s_{0},s_{0}^{\prime })$.
The pair $(M_{1}^{2},s_{0})$ describes the tetraquark channel, whereas $%
(M_{2}^{2},s_{0}^{\prime })$ corresponds to the $B_{c}^{\ast -}$ channel.

Requirements which should be satisfied by the auxiliary parameters $\mathbf{M%
}^{2}$ and $\mathbf{s}_{0}$ are universal for all SR computations and have
been discussed in the previous section. Thus, the windows for the Borel
parameters $M_{1}^{2}$ and $M_{2}^{2}$ should lead to a relatively stable
prediction for the form factor $g_{1}(q^{2})$. One has also to take into
account that $s_{0}$ and $s_{0}^{\prime }$ are restricted by the masses of
the excited hadrons in corresponding channels. Our computations demonstrate
that regions Eq.\ (\ref{eq:Wind1}) for parameters $(M_{1}^{2},s_{0})$ and
\begin{equation}
M_{2}^{2}\in \lbrack 6.5,7.5]~\mathrm{GeV}^{2},\ s_{0}^{\prime }\in \lbrack
46,47]~\mathrm{GeV}^{2},  \label{eq:Wind3}
\end{equation}%
which corresponds to the $B_{c}^{\ast -}$ channel, meet all constraints of
the sum rule analysis. Additionally, this choice for $(M_{1}^{2},s_{0})$
prevent a situation, when the mass and current coupling of the tetraquark $%
T_{\mathrm{PS}}$ entering to the sum rule Eq.\ (\ref{eq:SRCoup2}) 
generate new large uncertainties.

The sum rule method give credible results for the form factor $g_{1}(q^{2})$
in the Euclidean region $q^{2}<0$. But the strong coupling $g_{1}$ is
determined by $g_{1}(q^{2})$ at the mass shell $q^{2}=m_{B_{c}}^{2}$. To
evade this obstacle, it is convenient to introduce the function $%
g_{1}(Q^{2}) $ with $Q^{2}=-q^{2}$ and use it in following considerations.

\begin{figure}[h]
\includegraphics[width=8.5cm]{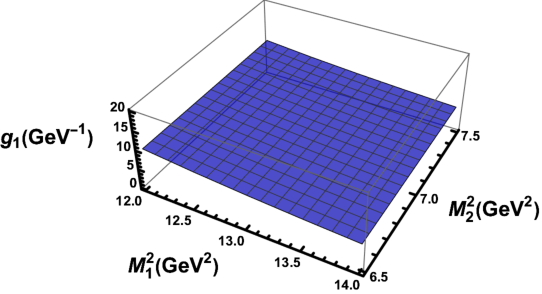}
\caption{Dependence of the form factor $g_{1}(4~\mathrm{GeV}^2)$ on the
Borel parameters $M_1^{2}$ and $M_2^{2}$ at middle values of $s_0$ and $%
s_{0}^{\prime}$. }
\label{fig:SCoupling}
\end{figure}

\begin{figure}[h]
\includegraphics[width=8.5cm]{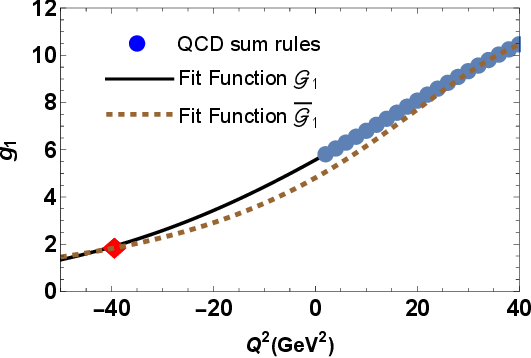}
\caption{QCD data and fit functions for the form factor $g_{1}(Q^{2})$. The
diamond is placed at $Q^{2}=-m_{B_{c}}^{2}$. }
\label{fig:Fit}
\end{figure}

In Fig.\ \ref{fig:SCoupling}, we depict the form factor $g_{1}(Q^{2})$ as a
function of the Borel parameters $(M_{1}^{2},M_{2}^{2})$ at fixed $Q^{2}=4~%
\mathrm{GeV}^{2}$ and $(s_{0},s_{0}^{\prime })$. It is seen that there is a
dependence of $g_{1}(Q^{2})$ on $M_{1}^{2}$ and $M_{2}^{2}$. The form factor
$g_{1}(Q^{2})$ at $Q^{2}=4~\mathrm{GeV}^{2}$ is equal to
\begin{equation}
g_{1}(4~\mathrm{GeV}^{2})=7.91\pm 1.89.
\end{equation}%
Here, uncertainties are generated by changes of the Borel parameters $%
(M_{1}^{2},M_{2}^{2})$. This effect and variations of $g_{1}(Q^{2})$ caused
by the continuum subtraction parameters are main sources of ambiguities in
sum rule computations. But, in our case they stay within limits $\pm 24\%$
which are acceptable for SR studies. We have varied $Q^{2}$ inside of the
region $Q^{2}=1-40~\mathrm{GeV}^{2}$ and collected information on $%
g_{1}(Q^{2})$: Obtained results are plotted in Fig.\ \ref{fig:Fit}.

As it has been emphasized above, the strong coupling $g_{1}$ should be
extracted at $q^{2}=m_{B_{c}}^{2}$, i.e., at $Q^{2}=-m_{B_{c}}^{2}$ where
the SR method does not work. Therefore, we introduce the fit functions $%
\mathcal{G}_{1}(Q^{2},m^{2})$ and $\overline{\mathcal{G}}_{1}(Q^{2},m^{2})$
that at momenta $Q^{2}>0$ give the same SR data, but can be extrapolated to
the region of negative $Q^{2}$. For these purposes, we use functions
\begin{equation}
\mathcal{G}_{i}(Q^{2},m^{2})=\mathcal{G}_{i}^{0}\mathrm{\exp }\left[
c_{i}^{1}\frac{Q^{2}}{m^{2}}+c_{i}^{2}\left( \frac{Q^{2}}{m^{2}}\right) ^{2}%
\right] ,  \label{eq:FitF}
\end{equation}%
and%
\begin{equation}
\overline{\mathcal{G}}_{i}(Q^{2},m^{2})=\frac{\overline{\mathcal{G}}_{i}^{0}%
}{\left( 1-\frac{Q^{2}}{m^{2}}\right) \left( 1-\sigma _{1}\frac{Q^{2}}{m^{2}}%
+\sigma _{2}\left( \frac{Q^{2}}{m^{2}}\right) ^{2}\right) },
\label{eq:FitF2}
\end{equation}%
where $\mathcal{G}_{i}^{0}$, $c_{i}^{1}$, and $c_{i}^{2},$ as well as $%
\overline{\mathcal{G}}_{i}^{0}$, $\sigma _{1}$ and $\sigma _{2}$ are fitted
constants.

Then, from analysis of the QCD data and Eqs.\ (\ref{eq:FitF}) and (\ref%
{eq:FitF2}), one can find the parameters $\mathcal{G}_{1}^{0}=5.57$, $%
c_{1}^{1}=3.67$, and $c_{1}^{2}=-4.15$ of the function $\mathcal{G}%
_{1}(Q^{2},m^{2})$, and $\overline{\mathcal{G}}_{i}^{0}=4.81$, $\sigma
_{1}=3.32$ and $\sigma _{2}=6.86$ of the function $\overline{\mathcal{G}}%
_{1}(Q^{2},m^{2})$. They are shown in Fig.\ \ref{fig:Fit}, in which one sees
very nice agreement of $\mathcal{G}_{1}(Q^{2},m^{2})$ and SR data. Having
used this function, we get $g_{1}$
\begin{equation}
g_{1}\equiv \mathcal{G}_{1}(-m_{B_{c}}^{2},m^{2})=1.92\pm 0.46.
\label{eq:g1}
\end{equation}%
Agreement of the function $\overline{\mathcal{G}}_{1}(Q^{2},m^{2})$ with SR
data is reasonable for high $Q^{2},$ whereas in low $Q^{2}$ region it
deviates from them. Nevertheless, at $Q^{2}=-m_{B_{c}}^{2}$ we find $%
\overline{\mathcal{G}}_{1}(-m_{B_{c}}^{2},m^{2})=1.84$ that is very close to
the result from Eq.\ (\ref{eq:g1}). The difference $0.08$ between these two
predictions is considerably smaller than theoretical errors in Eq.\ (\ref%
{eq:g1}). Therefore, throughout this work we employ Eq.\ (\ref{eq:FitF}) and
neglect small effects due to alternative extrapolating functions.

The width of the process $T_{\mathrm{PS}}\rightarrow B_{c}^{-}B_{c}^{\ast -}$
can be calculated by means of the expression%
\begin{equation}
\Gamma \left[ T_{\mathrm{PS}}\rightarrow B_{c}^{-}B_{c}^{\ast -}\right]
=g_{1}^{2}\frac{\lambda _{1}^{3}}{8\pi m_{1}^{2}},  \label{eq:PDw2}
\end{equation}%
where $\lambda _{1}=\lambda (m,m_{1},m_{B_{c}})$, and
\begin{equation}
\lambda (x,y,z)=\frac{\sqrt{%
x^{4}+y^{4}+z^{4}-2(x^{2}y^{2}+x^{2}z^{2}+y^{2}z^{2})}}{2x}.
\end{equation}%
As a result, we find
\begin{equation}
\Gamma \left[ T_{\mathrm{PS}}\rightarrow B_{c}^{-}B_{c}^{\ast -}\right]
=(19.8\pm 6.8)~\mathrm{MeV}.  \label{eq:DW1}
\end{equation}

\subsection{Decay $T_{\mathrm{PS}}\rightarrow B_{c}^{-}B_{c}^{-}(1^{3}P_{0})$%
}


The partial width of the process $T_{\mathrm{PS}}\rightarrow
B_{c}^{-}B_{c}^{-}(1^{3}P_{0})$ depends on the coupling $g_{2}$ at the
tetraquark-meson-meson vertex $T_{\mathrm{PS}}B_{c}^{-}B_{c}^{-}(1^{3}P_{0})$%
. To evaluate the coupling $g_{2}$, we start from the correlation function
\begin{eqnarray}
\Pi (p,p^{\prime }) &=&i^{2}\int d^{4}xd^{4}ye^{ip^{\prime
}y}e^{-ipx}\langle 0|\mathcal{T}\{J^{2}(y)  \notag \\
&&\times J^{B_{c}}(0)J^{\dag }(x)\}|0\rangle ,  \label{eq:CF3A}
\end{eqnarray}%
where $J^{2}(y)$ is the current for the scalar meson $B_{c}^{-}(1^{3}P_{0})$%
\begin{equation}
J^{2}(y)=\overline{c}_{i}(y)b_{i}(y).  \label{eq:CR3A}
\end{equation}

The physical side of SR is given by the expression%
\begin{eqnarray}
&&\Pi ^{\mathrm{Phys}}(p,p^{\prime })=\frac{\langle
0|J^{2}|B_{c}(1^{3}P_{0})(p^{\prime })\rangle }{p^{\prime 2}-m_{2}^{2}}\frac{%
\langle 0|J^{B_{c}}|B_{c}(q)\rangle }{q^{2}-m_{B_{c}}^{2}}  \notag \\
&&\times \langle B_{c}(1^{3}P_{0})(p^{\prime })B_{c}(q)|T_{\mathrm{PS}%
}(p)\rangle \frac{\langle T_{\mathrm{PS}}(p)|J^{\dag }|0\rangle }{p^{2}-m^{2}%
}  \notag \\
&&+\cdots ,  \label{eq:PhysSide4}
\end{eqnarray}%
with $m_{2}$ being the mass of the scalar meson $B_{c}^{-}(1^{3}P_{0})$. The
correlator $\Pi ^{\mathrm{Phys}}(p,p^{\prime })$ can be transformed to
standard form by employing the matrix elements
\begin{eqnarray}
&&\langle 0|J^{2}|B_{c}(1^{3}P_{0})\rangle =m_{2}f_{2},  \notag \\
&&\langle B_{c}(1^{3}P_{0})(p^{\prime })B_{c}(q)|T_{\mathrm{PS}}(p)\rangle
=g_{2}(q^{2})p\cdot p^{\prime }\text{.}  \label{eq:ME4}
\end{eqnarray}%
Here, $f_{2}$ is the decay constant of $B_{c}^{-}(1^{3}P_{0})$: Its
numerical value $f_{2}=(236\pm 17)~\mathrm{MeV}$ is borrowed from Ref.\ \cite%
{Wang:2024fwc}. After some calculations, we find that%
\begin{eqnarray}
&&\Pi ^{\mathrm{Phys}}(p,p^{\prime })=g_{2}(q^{2})\frac{|\Lambda
|f_{B_{c}}m_{B_{c}}^{2}m_{2}f_{2}}{(m_{b}+m_{c})\left( p^{2}-m^{2}\right) }
\notag \\
&&\times \frac{m^{2}+m_{2}^{2}-q^{2}}{2\left( p^{\prime 2}-m_{2}^{2}\right)
(q^{2}-m_{B_{c}}^{2})}+\cdots .  \label{eq:PhysSide4A}
\end{eqnarray}%
Because Eq.\ (\ref{eq:PhysSide4A}) consists of only a trivial Lorentz term,
its right-hand-side is equal to the invariant amplitude which we label $\Pi
_{2}^{\mathrm{Phys}}(p^{2},p^{\prime 2},q^{2})$.

The same correlator obtained in the QCD framework reads%
\begin{eqnarray}
&&\Pi ^{\mathrm{OPE}}(p,p^{\prime })=2i\int d^{4}xd^{4}ye^{ip^{\prime
}y}e^{-ipx}\left\{ \mathrm{Tr}\left[ S_{b}^{ia}(y-x)\right. \right.  \notag
\\
&&\left. \times \gamma _{5}\widetilde{S}_{b}^{jb}(-x)\gamma _{5}\widetilde{S}%
_{c}^{bj}(x)S_{c}^{ai}(x-y)\right]  \notag \\
&&\left. +\mathrm{Tr}\left[ S_{b}^{ia}(y-x)\gamma _{5}\widetilde{S}%
_{b}^{jb}(-x)\gamma _{5}\widetilde{S}_{c}^{aj}(x)S_{c}^{bi}(x-y)\right]
\right\} .  \notag \\
&&  \label{eq:QCDside3}
\end{eqnarray}%
Having denoted by $\Pi _{2}^{\mathrm{OPE}}(p^{2},p^{\prime 2},q^{2})$ the
relevant invariant amplitude, we get the sum rule for the form factor $%
g_{2}(q^{2})$
\begin{eqnarray}
&&g_{2}(q^{2})=\frac{2(m_{b}+m_{c})}{|\Lambda
|f_{B_{c}}m_{B_{c}}^{2}m_{2}f_{2}}\frac{q^{2}-m_{B_{c}}^{2}}{%
m^{2}+m_{2}^{2}-q^{2}}  \notag \\
&&\times e^{m^{2}/M_{1}^{2}}e^{m_{2}^{2}/M_{2}^{2}}\Pi _{2}(\mathbf{M}^{2},%
\mathbf{s}_{0},q^{2}).  \label{eq:SRCoup3}
\end{eqnarray}%
To carry out numerical computations, we fix the Borel and continuum
subtraction parameters. In the channel of the tetraquark $T_{\mathrm{PS}}$
they are the same as in Eq.\ (\ref{eq:Wind1}). In the channel of the $%
B_{c}^{-}(1^{3}P_{0})$ meson, we employ
\begin{equation}
M_{2}^{2}\in \lbrack 6.5,7.5]~\mathrm{GeV}^{2},\ s_{0}^{\prime }\in \lbrack
49,50]~\mathrm{GeV}^{2},  \label{eq:Wind4}
\end{equation}%
where $\sqrt{s_{0}^{\prime }}$ is limited by the mass $m[B_{c}(2^{3}P_{0})]=$
$7.122~\mathrm{GeV}$ of the first radially excited state $B_{c}(2^{3}P_{0})$
\cite{Godfrey:2004ya}.

Analysis performed in accordance with the scheme explained above leads to
the following predictions: For the extrapolating function $\mathcal{G}%
_{2}(Q^{2},m^{2})$, we get $\mathcal{G}_{2}^{0}=0.15\ \mathrm{GeV}^{-1}$, $%
c_{2}^{1}=0.99$, and $c_{2}^{2}=-0.25$. Then strong coupling $g_{2\text{ }}$%
extracted from $\mathcal{G}_{2}(Q^{2},m^{2})$ at the mass-shall $%
Q^{2}=-m_{B_{c}}^{2}$ is
\begin{equation}
g_{2}\equiv \mathcal{G}_{2}(-m_{B_{c}}^{2},m^{2})=(1.2\pm 0.3)\times
10^{-1}\ \mathrm{GeV}^{-1}.
\end{equation}%
The partial width of the decay $T_{\mathrm{PS}}\rightarrow
B_{c}^{-}B_{c}^{-}(1^{3}P_{0})$ is given by the formula%
\begin{equation}
\Gamma \left[ T_{\mathrm{PS}}\rightarrow B_{c}^{-}B_{c}^{-}(1^{3}P_{0})%
\right] =g_{2}^{2}\frac{m_{2}^{2}\lambda _{2}}{8\pi }\left( 1+\frac{\lambda
_{2}^{2}}{m_{2}^{2}}\right) ,  \label{eq:PDW2}
\end{equation}%
where $\lambda _{2}=\lambda (m,m_{2},m_{B_{c}})$. Our computations yield
\begin{equation}
\Gamma \left[ T_{\mathrm{PS}}\rightarrow B_{c}^{-}B_{c}^{-}(1^{3}P_{0})%
\right] =(21.8\pm 7.8)~\mathrm{MeV}.  \label{eq:DW2}
\end{equation}


\subsection{Decay $T_{\mathrm{PS}}\rightarrow B_{c}^{\ast
-}B_{c}^{-}(1^{1}P_{1})$}


The process $T_{\mathrm{PS}}\rightarrow B_{c}^{\ast -}B_{c}^{-}(1^{1}P_{1})$
can be considered in accordance with techniques presented in previous
subsections. Differences here are connected with spin-parities of the
involved particles and corresponding matrix elements. In the case under
analysis, we should consider the correlation function
\begin{eqnarray}
\Pi _{\mu \nu }(p,p^{\prime }) &=&i^{2}\int d^{4}xd^{4}ye^{ip^{\prime
}y}e^{-ipx}\langle 0|\mathcal{T}\{J_{\mu }^{3}(y)  \notag \\
&&\times J_{\nu }^{1}(0)J^{\dag }(x)\}|0\rangle .  \label{eq:CF4}
\end{eqnarray}%
The interpolating current $J_{\mu }^{3}(y)$ for the axial-vector meson $%
B_{c}^{-}(1^{1}P_{1})$ is
\begin{equation}
J_{\mu }^{3}(y)=\overline{c}_{i}(y)\gamma _{\mu }\gamma _{5}b_{i}(y).
\label{eq:C4}
\end{equation}%
The matrix element of $B_{c}^{-}(1^{1}P_{1})$, as well as the vertex which
are necessary for our purposes have the forms
\begin{eqnarray}
&&\langle 0|J_{\mu }^{3}|B_{c}(1^{1}P_{1})(p^{\prime },\epsilon )\rangle
=m_{3}f_{3}\epsilon _{\mu }(p^{\prime }),  \notag \\
&&\langle B_{c}(1^{1}P_{1})(p^{\prime },\epsilon )B_{c}^{\ast
}(q,\varepsilon )|T_{\mathrm{PS}}(p)\rangle =g_{3}(q^{2})\left[ \left(
p\cdot p^{\prime }\right) \right.  \notag \\
&&\left. \times \left( \epsilon ^{\ast }\cdot \varepsilon ^{\ast }\right)
-\left( p^{\prime }\cdot \varepsilon ^{\ast }\right) \left( p\cdot \epsilon
^{\ast }\right) \right] ,  \label{eq:ME5}
\end{eqnarray}%
where $m_{3}$, $f_{3}$, and $\epsilon _{\mu }(p^{\prime })$ are the mass,
decay constant and polarization vector of the meson $B_{c}^{-}(1^{1}P_{1})$.
The mass of $B_{c}^{-}(1^{1}P_{1})$ has been presented above, whereas for
its decay constant, we utilize $f_{3}=(373\pm 25)~\mathrm{MeV}$ from Ref.\
\cite{Wang:2012kw}.

The correlator $\Pi _{\mu \nu }(p,p^{\prime })$ obtained using the physical
parameters of particles is equal to
\begin{eqnarray}
&&\Pi _{\mu \nu }^{\mathrm{Phys}}(p,p^{\prime })=g_{3}(q^{2})\frac{|\Lambda
|f_{1}f_{3}}{2m_{1}m_{3}\left( p^{2}-m^{2}\right) \left( p^{\prime
2}-m_{3}^{2}\right) }  \notag \\
&&\times \frac{1}{(q^{2}-m_{1}^{2})}\left\{ \left(
m^{2}-m_{3}^{2}-q^{2}\right) m_{3}^{2}\left( m_{1}^{2}g_{\mu \nu }-2p_{\mu
}p_{\nu }\right) \right.  \notag \\
&&-\left[ m^{4}-m_{3}^{4}+2m^{2}m_{1}^{2}-q^{2}(2m^{2}-2m_{1}^{2}-q^{2})%
\right] p_{\mu }^{\prime }p_{\nu }^{\prime }  \notag \\
&&+2m_{3}^{2}(m^{2}-m_{3}^{2}-q^{2}+m_{1}^{2})p_{\mu }p_{\nu }^{\prime }+
\left[ m^{4}-m_{3}^{4}\right.  \notag \\
&&\left. \left. -q^{2}(2m^{2}+q^{2})\right] p_{\mu }^{\prime }p_{\nu
}\right\} +\cdots .  \label{eq:PhysSide6}
\end{eqnarray}

In terms of quark propagators $\Pi _{\mu \nu }(p,p^{\prime })$ reads%
\begin{eqnarray}
&&\Pi _{\mu \nu }^{\mathrm{OPE}}(p,p^{\prime })=-2i\int
d^{4}xd^{4}ye^{ip^{\prime }y}e^{-ipx}\left\{ \mathrm{Tr}\left[ \gamma _{\mu
}\gamma _{5}\right. \right.  \notag \\
&&\left. \times S_{b}^{ia}(y-x)\gamma _{5}\widetilde{S}_{b}^{jb}(-x)\gamma
_{\nu }\widetilde{S}_{c}^{bj}(x)S_{c}^{ai}(x-y)\right]  \notag \\
&&\left. +\mathrm{Tr}\left[ \gamma _{\mu }\gamma _{5}S_{b}^{ia}(y-x)\gamma
_{5}\widetilde{S}_{b}^{jb}(-x)\gamma _{\nu }\widetilde{S}%
_{c}^{aj}(x)S_{c}^{bi}(x-y)\right] \right\} .  \notag \\
&&  \label{eq:QCDSide4}
\end{eqnarray}%
The SR for the form factor $g_{3}(q^{2})$ is obtained using the invariant
amplitudes corresponding to structures $\sim g_{\mu \nu }$ in both $\Pi
_{\mu \nu }^{\mathrm{Phys}}(p,p^{\prime })$ and $\Pi _{\mu \nu }^{\mathrm{OPE%
}}(p,p^{\prime })$. The numerical computations have been carried out in
accordance with the scheme described in this section, where for $M_{2}^{2}\ $%
and $s_{0}^{\prime }$ in the $B_{c}^{-}(1^{1}P_{1})$ channel, we use the
following values
\begin{equation}
M_{2}^{2}\in \lbrack 6.5,7.5]~\mathrm{GeV}^{2},\ s_{0}^{\prime }\in \lbrack
50,51]~\mathrm{GeV}^{2}.  \label{eq:Wind5}
\end{equation}%
These parameters are determined by the mass $7.15~\mathrm{GeV}$ of the
radially excited meson $B_{c}^{-}(2^{1}P_{1})$, and the requirement $%
s_{0}^{\prime }<51.12~\mathrm{GeV}^{2}$.

Our analysis leads to the fit function $\mathcal{G}_{3}(Q^{2},m^{2})$ with
parameters $\mathcal{G}_{3}^{0}=0.32\ \mathrm{GeV}^{-1}$, $c_{3}^{1}=1.54$,
and $c_{3}^{2}=-5.44$. Then, the strong coupling $g_{3}$ is
\begin{equation}
g_{3}\equiv \mathcal{G}_{3}(-m_{1}^{2},m^{2})=(1.6\pm 0.4)\times 10^{-1}\
\mathrm{GeV}^{-1}.
\end{equation}%
The width of the decay $T_{\mathrm{PS}}\rightarrow B_{c}^{\ast
-}B_{c}^{-}(1^{1}P_{1})$ is equal to
\begin{equation}
\Gamma \left[ T_{\mathrm{PS}}\rightarrow B_{c}^{\ast -}B_{c}^{-}(1^{1}P_{1})%
\right] =g_{3}^{2}\frac{\lambda _{3}}{8\pi }|\mathcal{M}|^{2},
\label{eq:PDW3}
\end{equation}%
\newline
where
\begin{eqnarray}
&&|\mathcal{M}|^{2}=m^{2}\left[ 1-\zeta \left( 1-\frac{1}{2}\zeta \right) %
\right]  \notag \\
&&+m_{3}^{2}\left( \frac{3}{2}-\frac{1}{2}\eta -\frac{3}{4}\zeta +\frac{1}{%
4\zeta }+\frac{1}{4\zeta }\eta \right) ,
\end{eqnarray}%
$\lambda _{3}=\lambda (m,m_{3},m_{1})$, $\zeta =m_{1}^{2}/m^{2}$ and $\eta
=m_{3}^{2}/m_{1}^{2}$. Our calculations give
\begin{equation}
\Gamma \left[ T_{\mathrm{PS}}\rightarrow B_{c}^{\ast -}B_{c}^{-}(1^{1}P_{1})%
\right] =(22.1\pm 7.8)~\mathrm{MeV}.  \label{eq:DW3}
\end{equation}%
With these results at hand it is easy to estimate the full width of the
tetraquark $T_{\mathrm{PS}}$%
\begin{equation}
\Gamma _{\mathrm{PS}}=(63.7\pm 13.0)~\mathrm{MeV}.
\end{equation}


\section{Width of the state $T_{\mathrm{V}}$}

\label{sec:VWidths}

The vector tetraquark has the mass $\widetilde{m}=(13.15\pm 0.10)~\mathrm{GeV%
}$, and hence can dissociate to meson pairs $2B_{c}^{-}$, $2B_{c}^{\ast -}$%
and $B_{c}^{-}B_{c}^{-}(1^{1}P_{1})$. In fact, the thresholds for these
processes $12.55~\mathrm{GeV}$, $12.676~\mathrm{GeV}$ and $13.025~\mathrm{GeV%
}$ are below the mass of $T_{\mathrm{V}}$.


\subsection{$T_{\mathrm{V}}\rightarrow B_{c}^{\ast -}B_{c}^{\ast -}$}


To consider the process $T_{\mathrm{V}}\rightarrow B_{c}^{\ast -}B_{c}^{\ast
-}$, we begin from analysis of the correlation function
\begin{eqnarray}
\Pi _{\mu \delta \nu }(p,p^{\prime }) &=&i^{2}\int d^{4}xd^{4}ye^{ip^{\prime
}y}e^{-ipx}\langle 0|\mathcal{T}\{J_{\mu }^{1}(y)  \notag \\
&&\times J_{\delta }^{1}(0)J_{\nu }^{\dag }(x)\}|0\rangle .  \label{eq:CF7}
\end{eqnarray}%
First, we express $\Pi _{\mu \delta \nu }(p,p^{\prime })$ by means of
physical parameters of the particles $T_{\mathrm{V}}$ and $B_{c}^{\ast -}$,
and write it in the form
\begin{eqnarray}
&&\Pi _{\mu \delta \nu }^{\mathrm{Phys}}(p,p^{\prime })=\frac{\langle
0|J_{\mu }^{1}|B_{c}^{\ast }(p^{\prime },\varepsilon (p^{\prime }))\rangle }{%
p^{\prime 2}-m_{1}^{2}}\frac{\langle 0|J_{\delta }^{1}|B_{c}^{\ast
}(q,\varepsilon (q))\rangle }{q^{2}-m_{1}^{2}}  \notag \\
&&\times \langle B_{c}^{\ast }(p^{\prime },\varepsilon (p^{\prime
}))B_{c}^{\ast }(q,\varepsilon (q))|T_{\mathrm{V}}(p,\epsilon (p))\rangle
\frac{\langle T_{\mathrm{V}}(p,\epsilon (p))|J_{\nu }^{\dag }|0\rangle }{%
p^{2}-\widetilde{m}^{2}}  \notag \\
&&+\cdots .  \label{eq:PhysSide2}
\end{eqnarray}%
The matrix elements of the particles $B_{c}^{\ast -}$ and $T_{\mathrm{V}}$
have been introduced in previous sections. The vertex $T_{\mathrm{V}%
}B_{c}^{\ast -}B_{c}^{\ast -}$ is characterized by the form factor $%
G_{1}(q^{2})$ and has the following decomposition
\begin{eqnarray}
&&\langle B_{c}^{\ast }(p^{\prime },\varepsilon (p^{\prime }))B_{c}^{\ast
}(q,\varepsilon (q))|T_{\mathrm{V}}(p,\epsilon (p))\rangle =G_{1}(q^{2})
\notag \\
&&\times \lbrack (q-p^{\prime })_{\gamma }g_{\alpha \beta }-(p+q)_{\alpha
}g_{\gamma \beta }+(p+q)_{\beta }g_{\gamma \alpha }]  \notag \\
&&\times \epsilon ^{\gamma }(p)\varepsilon ^{\ast \alpha }(p^{\prime
})\varepsilon ^{\ast \beta }(q).  \label{eq:Mel}
\end{eqnarray}%
The correlation function $\Pi _{\mu \delta \nu }^{\mathrm{Phys}}(p,p^{\prime
})$ through parameters of the tetraquark $T_{\mathrm{V}}$ and meson $%
B_{c}^{\ast -}$ reads
\begin{eqnarray}
&&\Pi _{\mu \delta \nu }^{\mathrm{Phys}}(p,p^{\prime })=\frac{G_{1}(q^{2})|%
\widetilde{\Lambda }|f_{1}^{2}m_{1}^{2}}{(p^{2}-\widetilde{m}^{2})(p^{\prime
2}-m_{1}^{2})(q^{2}-m_{1}^{2})}  \notag \\
&&\times \left[ \frac{\widetilde{m}^{2}+m_{1}^{2}-q^{2}}{2m_{1}^{4}}p_{\nu
}p_{\delta }p_{\mu }^{\prime }-\frac{\widetilde{m}^{2}}{m_{1}^{2}}g_{\mu \nu
}p_{\delta }^{\prime }\right.  \notag \\
&&-\frac{\widetilde{m}^{2}-2m_{1}^{2}}{m_{1}^{4}}\left( p_{\delta }p_{\nu
}^{\prime }-p_{\delta }^{\prime }p_{\nu }^{\prime }\right) p_{\mu }^{\prime
}+2g_{\mu \delta }p_{\nu }^{\prime }+2g_{\nu \delta }p_{\mu }  \notag \\
&&\left. -\left( \widetilde{m}^{2}+m_{1}^{2}-q^{2}\right) \left( \frac{%
g_{\mu \delta }p_{\nu }}{\widetilde{m}^{2}}+\frac{g_{\nu \delta }p_{\mu
}^{\prime }}{m_{1}^{2}}\right) ...\right] ,  \label{eq:PhysSide2a}
\end{eqnarray}%
where the ellipses denote other terms in the correlator.

The same correlation function obtained using the quark propagators has the
following form
\begin{eqnarray}
&&\Pi _{\mu \delta \nu }^{\mathrm{OPE}}(p,p^{\prime })=2i^{2}\int
d^{4}xd^{4}ye^{ip^{\prime }y}e^{-ipx}\left\{ \mathrm{Tr}\left[ \gamma _{\mu
}S_{b}^{ia}(y-x)\right. \right.  \notag \\
&&\left. \times \gamma _{5}\widetilde{S}_{b}^{jb}(-x)\gamma _{\delta }%
\widetilde{S}_{c}^{bj}(x)\gamma _{\nu }\gamma _{5}S_{c}^{ai}(x-y)\right]
\notag \\
&&\left. +\mathrm{Tr}\left[ \gamma _{\mu }S_{b}^{ia}(y-x)\gamma _{5}%
\widetilde{S}_{b}^{jb}(-x)\gamma _{\delta }\widetilde{S}_{c}^{aj}(x)\gamma
_{\nu }\gamma _{5}S_{c}^{bi}(x-y)\right] \right\} .  \notag \\
&&  \label{eq:QCDside2a}
\end{eqnarray}%
Having used amplitudes corresponding to the structures $\sim p_{\nu
}p_{\delta }p_{\mu }^{\prime }$ in both the physical and QCD expressions for
the correlation function, we derive the SR for the form factor $G_{1}(q^{2})
$.

\begin{figure}[h]
\includegraphics[width=8.5cm]{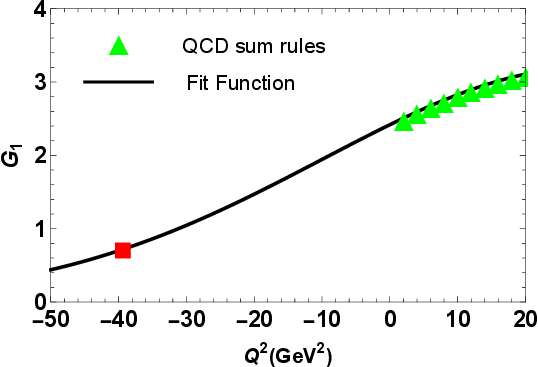}
\caption{Results of the sum rule computations and the extrapolating function
for the form factor $G_{1}(Q^{2})$. The red square is placed at the point $%
Q^{2}=-m_{1}^{2}$. }
\label{fig:Fit1}
\end{figure}

At the mass shell $q^{2}=m_{1}^{2}$ the strong coupling $G_{1}$ is equal to
\begin{equation}
G_{1}\equiv \widetilde{\mathcal{G}}_{1}(-m_{1}^{2},\widetilde{m}%
^{2})=(6.9\pm 1.6)\times 10^{-1}\ ,
\end{equation}%
which is found by means of the function $\widetilde{\mathcal{G}}_{1}$ with $%
\widetilde{\mathcal{G}}_{1}^{0}=2.42$, $\widetilde{c}_{1}^{1}=3.23$, and $%
\widetilde{c}_{1}^{2}=-9.20$. The results of the sum rule computations in
the region $Q^{2}=1-20~\mathrm{GeV}^{2}$, as well as the extrapolating
function $\widetilde{\mathcal{G}}_{1}$ are depictured in Fig.\ \ref{fig:Fit1}%
.

The width of the decay $T_{\mathrm{V}}\rightarrow B_{c}^{\ast -}B_{c}^{\ast
-}$ is given by the following expression
\begin{equation}
\Gamma \left[ T_{\mathrm{V}}\rightarrow B_{c}^{\ast -}B_{c}^{\ast -}\right]
=G_{1}^{2}\frac{\widetilde{\lambda }_{1}}{48\pi }\left( \frac{1}{4\widetilde{%
\zeta }^{2}}-\frac{4}{\widetilde{\zeta }}-12\widetilde{\zeta }-17\right) ,
\end{equation}%
where $\widetilde{\lambda }_{1}=\lambda (\widetilde{m},m_{1},m_{1})$, and $%
\widetilde{\zeta }=m_{1}^{2}/\widetilde{m}^{2}$. Then, for the width of this
channel we find
\begin{equation}
\Gamma \left[ T_{\mathrm{V}}\rightarrow B_{c}^{\ast -}B_{c}^{\ast -}\right]
=(11.5\pm 3.9)~\mathrm{MeV}.
\end{equation}

\subsection{$T_{\mathrm{V}}\rightarrow B_{c}^{-}B_{c}^{-}$ and $T_{\mathrm{V}%
}\rightarrow B_{c}^{-}B_{c}^{-}(1^{1}P_{1})$}


These processes can be explored by applying the methods described in a
rather detailed form in previous parts of the article. Therefore, there is
not a necessity to write down repeatedly similar expressions for the
correlation functions using different interpolating currents. Instead, it is
convenient to present the new matrix elements and formulas for the
extrapolating functions, partial decay widths typical for these processes.

The main ingredients in SR analysis are matrix elements of the particles
involved into the decay processes. In the case of the decay $T_{\mathrm{V}%
}\rightarrow B_{c}^{-}B_{c}^{-}$ matrix element specific for this channel is
\begin{equation}
\langle B_{c}(p^{\prime })B_{c}(q)|T_{\mathrm{V}}(p,\epsilon (p))\rangle
=G_{2}(q^{2})\epsilon (p)\cdot p^{\prime },
\end{equation}%
while other matrix elements have been introduced above. For the decay $T_{%
\mathrm{V}}\rightarrow B_{c}^{-}B_{c}^{-}(1^{1}P_{1})$ relevant vertex is
defined in accordance with the expression%
\begin{eqnarray}
&&\langle B_{c}(1^{1}P_{1})(p^{\prime },\varepsilon (p^{\prime
}))B_{c}(q)|T_{\mathrm{V}}(p,\epsilon (p))\rangle  \notag \\
&=&G_{3}(q^{2})\left[ \left( q\cdot p^{\prime }\right) \left( \epsilon \cdot
\varepsilon ^{\ast }\right) -\left( q\cdot \varepsilon ^{\ast }\right)
\left( p^{\prime }\cdot \epsilon \right) \right] .
\end{eqnarray}%
The form factors $G_{2}(q^{2})$ and $G_{3}(q^{2})$ determine the strong
interactions of particles at the vertices $T_{\mathrm{V}}B_{c}^{-}B_{c}^{-}$
and $T_{\mathrm{V}}B_{c}^{-}B_{c}^{-}(1^{1}P_{1})$, respectively.

The coupling $G_{2}$ is extracted at the mass shell $q^{2}=m_{B_{c}}^{2}$
and equals to
\begin{equation}
G_{2}\equiv \widetilde{\mathcal{G}}_{2}(-m_{B_{c}}^{2},\widetilde{m}%
^{2})=(7.4\pm 1.8).
\end{equation}%
It has been found using the extrapolation function $\widetilde{\mathcal{G}}%
_{2}$ with parameters $\widetilde{\mathcal{G}}_{2}^{0}=13.78$, $\widetilde{c}%
_{2}^{1}=1.76$, and $\widetilde{c}_{2}^{2}=-4.14$. The partial width of this
decay is equal to
\begin{equation}
\Gamma \left[ T_{\mathrm{V}}\rightarrow B_{c}^{-}B_{c}^{-}\right] =G_{2}^{2}%
\frac{\widetilde{\lambda }_{2}}{192\pi }\left( 1-\frac{4m_{B_{c}}^{2}}{%
\widetilde{m}^{2}}\right) ,
\end{equation}%
where $\widetilde{\lambda }_{2}=\lambda (\widetilde{m},m_{B_{c}},m_{B_{c}})$%
, and amounts to
\begin{equation}
\Gamma \left[ T_{\mathrm{V}}\rightarrow B_{c}^{-}B_{c}^{-}\right] =(16.1\pm
5.4)~\mathrm{MeV}.
\end{equation}

The strong coupling $G_{3}$ has been determined at $q^{2}=m_{B_{c}}^{2}$ by
applying the fit function $\widetilde{\mathcal{G}}_{3}$ with $\widetilde{%
\mathcal{G}}_{3}^{0}=0.66\ \mathrm{GeV}^{-1}$, $\widetilde{c}_{3}^{1}=4.85$,
and $\widetilde{c}_{3}^{2}=-9.44$. Our computations yield
\begin{equation}
G_{3}\equiv \widetilde{\mathcal{G}}_{3}(-m_{B_{c}}^{2},\widetilde{m}%
^{2})=(1.4\pm 0.3)\times 10^{-1}\ \mathrm{GeV}^{-1}.
\end{equation}%
The width of the decay $T_{\mathrm{V}}\rightarrow
B_{c}^{-}B_{c}^{-}(1^{1}P_{1})$ equals to
\begin{equation}
\Gamma \left[ T_{\mathrm{V}}\rightarrow B_{c}^{-}B_{c}^{-}(1^{1}P_{1})\right]
=(25.9\pm 7.8)~\mathrm{MeV}.
\end{equation}%
Then, the full width of the vector tetraquark $T_{\mathrm{V}}$ amounts to
\begin{equation}
\Gamma _{\mathrm{V}}=(53.5\pm 10.3)~\mathrm{MeV}.
\end{equation}


\section{Summing up}

\label{sec:Conc}

\begin{table}[tbp]
\begin{tabular}{|c|c|c|c|}
\hline\hline
$i$ & Channels & $g_{i}[G_{i}]\times10~(\mathrm{GeV}^{-1})$ & $\Gamma_{i}~(%
\mathrm{MeV})$ \\ \hline
$1$ & $T_{\mathrm{PS}}\to B_{c}^{-}B_{c}^{\ast -}$ & $19.2 \pm 4.6$ & $19.8
\pm 6.8$ \\
$2$ & $T_{\mathrm{PS}}\to B_{c}^{-}B_{c}^{-}(1^3P_{0})$ & $1.2 \pm 0.3$ & $%
21.8 \pm 7.8$ \\
$3$ & $T_{\mathrm{PS}} \to B_{c}^{\ast-}B_{c}^{-}(1^1P_{1})$ & $1.6 \pm 0.4$
& $22.1 \pm 7.8 $ \\ \hline
$1$ & $T_{\mathrm{V}}\to B_{c}^{\ast -}B_{c}^{\ast -}$ & $6.9 \pm 1.6$ & $%
11.5 \pm 3.9$ \\
$2$ & $T_{\mathrm{V}}\to B_{c}^{-}B_{c}^{-}$ & $74 \pm 18$ & $16.1 \pm 5.4$
\\
$3$ & $T_{\mathrm{V}} \to B_{c}^{-}B_{c}^{-}(1^1P_{1})$ & $1.4 \pm 0.3$ & $%
25.9 \pm 7.8 $ \\ \hline\hline
\end{tabular}%
\caption{Decay channels of the tetraquarks $T_{\mathrm{PS}}$ and $T_{\mathrm{%
V}}$, strong couplings $g_{i}$, $G_{i}$, and partial widths $\Gamma_{i}$.
The couplings $g_1$, $G_1$ and $G_2$ are dimensionless.}
\label{tab:Channels}
\end{table}

In this paper, we have explored the pseudoscalar $T_{\mathrm{PS}}$ and
vector $T_{\mathrm{V}}$ tetraquarks $bb\overline{c}\overline{c}$ by treating
these particles as diquark-antidiquark systems with structures $C\gamma
_{5}\otimes C$ and $C\gamma _{5}\otimes \gamma _{\mu }\gamma _{5}C$,
respectively. Predictions for their masses $(13.092\pm 0.095)~\mathrm{GeV}$
and $(13.15\pm 0.10)~\mathrm{GeV}$ confirm that they can easily dissociate
to $B_{c}^{(\ast )}B_{c}^{(\ast )}$ meson pairs.

We have calculated also the full widths of the tetraquarks $T_{\mathrm{PS}}$
and $T_{\mathrm{V}}$ (for details, see Table\ \ref{tab:Channels}). The full
width $\Gamma _{\mathrm{PS}}=(63.7\pm 13.0)~\mathrm{MeV}$ of the tetraquark $%
T_{\mathrm{PS}}$ is a sum of the partial widths of the processes $T_{\mathrm{%
PS}}\rightarrow B_{c}^{-}B_{c}^{\ast -}$, $B_{c}^{-}B_{c}^{-}(1^{3}P_{0})$
and $B_{c}^{\ast -}B_{c}^{-}(1^{1}P_{1})$. These three decay channels
contribute almost on the equal footing to the full width of $T_{\mathrm{PS}}$%
. The kinematically allowed channels of $T_{\mathrm{V}}$ are decays to meson
pairs $2B_{c}^{\ast -}$, $2B_{c}^{-}$ and $B_{c}^{-}B_{c}^{-}(1^{1}P_{1})$.
In the case of the vector state $T_{\mathrm{V}}$ dissociation to mesons $%
B_{c}^{-}B_{c}^{-}(1^{1}P_{1})$ is evidently its dominant decay mode. The
full width of $T_{\mathrm{V}}$ saturated by these processes is equal to $%
\Gamma _{\mathrm{V}}=(53.5\pm 10.3)~\mathrm{MeV} $. Founded on these
predictions, we classify $T_{\mathrm{PS}}$ and $T_{\mathrm{V}}$ as
tetraquarks with relatively moderate widths.

The parameters of the pseudoscalar and vector tetraquarks $T_{\mathrm{PS}}$
and $T_{\mathrm{V}}$ were calculated in some papers collected in Table\ \ref%
{tab:Theory}. In the framework of SR method these particles were explored in
Ref.\ \cite{Wang:2021taf} predictions of which are considerably lower than
our results, whereas a mass gap between them is almost the same. The
relativistic quark model leads to estimations \cite{Galkin:2023wox}, which
are compatible with our predictions provided one takes into account
uncertainties of SR computations.

\begin{table}[t]
\begin{tabular}{|c|c|c|}
\hline\hline
Works & $T_{\mathrm{PS}}$ & $T_{\mathrm{V}}$ \\ \hline
This work & $13.092\pm 0.095$ & $13.15\pm 0.10$ \\
\cite{Galkin:2023wox} & $13.106$ & $13.103-13.111$ \\
\cite{Wang:2021taf} & $12.72_{-0.19}^{+0.22}$ & $12.77_{-0.19}^{+0.24}$ \\
\hline\hline
\end{tabular}%
\caption{Theoretical predictions for the masses of the pseudoscalar and
vector states $bb\overline{c}\overline{c}$ obtained in different articles.
All masses are presented in $\mathrm{GeV}$ units.}
\label{tab:Theory}
\end{table}

The present analysis completes an intermediate stage in our investigations
of the exotic mesons with contents $bb\overline{c}\overline{c}$ and
spin-parities $J^{\mathrm{P}}=0^{+} $, $0^{-}$, $1^{+}$, $1^{-}$ and $2^{+}$%
. Having used QCD sum rule method and the diquark-antidiquark model, we
calculated masses and widths of these tetraquarks. All of them were found
unstable against two-meson decay channels. In this aspect, we agree with
main part of previous publications though there are controversies with a few
articles discussed in Refs.\ \cite{Agaev:2023tzi,Agaev:2024pej,Agaev:2024pil}%
.

Our studies do not embrace all diquark-antidiquark states with different
inner organizations. Moreover, the fully heavy exotic mesons may have
hadronic molecule structure parameters of which, especially their widths,
require detailed analysis. The physical resonances $bb\overline{c}\overline{c%
}$ may be superpositions of different diquark-antidiquark and/or molecule
states with the same quantum numbers. All these problems should be addressed
in the framework of different methods. Experimental information about
properties of the four-quark mesons $bb\overline{c}\overline{c}$/$cc%
\overline{b}\overline{b}$ will have crucial importance for related
theoretical studies.

\begin{widetext}
\appendix*

\section{ Heavy quark propagator and correlation functions}

\renewcommand{\theequation}{\Alph{section}.\arabic{equation}} \label{sec:App}


In this article, for the heavy quark propagator $S_{Q}^{ab}(x)$ ($Q=c,\ b$),
we employ the following expression
\begin{eqnarray}
&&S_{Q}^{ab}(x)=i\int \frac{d^{4}k}{(2\pi )^{4}}e^{-ikx}\Bigg \{\frac{\delta
_{ab}\left( {\slashed k}+m_{Q}\right) }{k^{2}-m_{Q}^{2}}-\frac{%
g_{s}G_{ab}^{\alpha \beta }}{4}\frac{\sigma _{\alpha \beta }\left( {\slashed %
k}+m_{Q}\right) +\left( {\slashed k}+m_{Q}\right) \sigma _{\alpha \beta }}{%
(k^{2}-m_{Q}^{2})^{2}}  \notag \\
&&+\frac{g_{s}^{2}G^{2}}{12}\delta _{ab}m_{Q}\frac{k^{2}+m_{Q}{\slashed k}}{%
(k^{2}-m_{Q}^{2})^{4}}+\cdots \Bigg \}.
\end{eqnarray}%
Here, we have introduced the notations
\begin{equation}
G_{ab}^{\alpha \beta }\equiv G_{A}^{\alpha \beta }\lambda _{ab}^{A}/2,\ \
G^{2}=G_{\alpha \beta }^{A}G_{A}^{\alpha \beta },\
\end{equation}%
with $G_{A}^{\alpha \beta }$ being the gluon field-strength tensor, and $%
\lambda ^{A}$--Gell-Mann matrices. The index $A$ varies in the range $1-8$.

This Appendix also contains expressions of correlation functions, which are
employed to calculate masses of the pseudoscalar and vector tetraquarks $bb%
\overline{c}\overline{c}$. The correlation function $\Pi (M^{2},s_{0})$
which enters to SRs for the mass of the $\mathrm{PS}$ meson has the
following form as presented in Eq.\ (\ref{eq:CorrF}):
\begin{equation}
\Pi (M^{2},s_{0})=\int_{4\mathcal{M}^{2}}^{s_{0}}ds\rho ^{\mathrm{OPE}%
}(s)e^{-s/M^{2}}+\Pi (M^{2}).  \label{eq:InvAmp}
\end{equation}%
\ In the case of the vector particle, the relevant correlator $\widetilde{\Pi
}(M^{2},s_{0})$ is given by the same expression but depends on $\widetilde{%
\rho }^{\mathrm{OPE}}(s)$ and $\widetilde{\Pi }(M^{2})$.

The spectral density $\rho ^{\mathrm{OPE}}(s)$ is determined as an imaginary
part of the function $\Pi ^{\mathrm{OPE}}(p)$ and is a sum of perturbative $%
\rho ^{\mathrm{pert.}}(s)$ and nonperturbative $\rho ^{\mathrm{Dim4}}(s)$
terms, which are given by the general expression%
\begin{equation}
\rho (s)=\int_{0}^{1}d\alpha \int_{0}^{1-\alpha }d\beta \int_{0}^{1-\alpha
-\beta }d\gamma \rho (s,\alpha ,\beta ,\gamma ),  \label{eq:A4}
\end{equation}%
with $\alpha $, $\beta $, and $\gamma $ being the Feynman parameters. The
second component in Eq.\ (\ref{eq:InvAmp}) is Borel transformation of some of
dimension-$4$ terms obtained directly from $\Pi ^{\mathrm{OPE}}(p^{2})$ and
which are not covered by the spectral density. The function $\Pi (M^{2})$ is
also determined by Eq.\ (\ref{eq:A4})-type formula with the integrand $\Pi
(M^{2},\alpha ,\beta ,\gamma )$.

In the case of the pseudoscalar particle for the perturbative component $%
\rho ^{\mathrm{pert.}}(s,\alpha ,\beta ,\gamma )$ we get:
\begin{eqnarray}
&&\rho ^{\mathrm{pert.}}(s,\alpha ,\beta ,\gamma )=\frac{N^{2}\Theta ({N)}}{%
128C^{4}B^{4}\pi ^{6}}\left\{ 4CB^{2}N\left[ B^{2}m_{b}^{2}\alpha \beta
+L_{1}\gamma (G^{2}m_{c}^{2}+3L_{1}s\alpha ^{2}\beta ^{2}\gamma )\right]
-3L_{1}B^{4}N^{2}\alpha \beta \gamma \right.  \notag \\
&&\left. -6C^{2}(B^{2}m_{c}^{2}+L_{1}s\alpha ^{2}\beta ^{2}\gamma
)(B^{2}m_{b}^{2}+L_{1}^{2}s\alpha \beta \gamma ^{2})\right\} ,
\end{eqnarray}%
where $\Theta ({z)}$ is the Unit Step function. Here%
\begin{equation}
N=-C\left[ s\alpha \beta \gamma L_{1}+A(m_{c}^{2}L_{2}-m_{b}^{2}(\alpha
+\beta ))\right] /A^{2},
\end{equation}%
and
\begin{eqnarray}
&&A=\beta \gamma L_{3}+\alpha ^{2}(\beta +\gamma )+\alpha \left[ \beta
(\beta +2\gamma -1)+\gamma (\gamma -1)\right] ,\ B=L_{3}C+\alpha ^{2}(\beta
+\gamma ),\   \notag \\
\text{ } &&C=\alpha \beta +\alpha \gamma +\beta \gamma ,\ G=L_{2}C+\gamma
^{2}(\beta +\alpha ),
\end{eqnarray}%
We also use the notations

\begin{equation}
L_{1}=\alpha +\beta +\gamma -1,\ L_{2}=\alpha +\beta -1,~L_{3}=\beta +\gamma
-1.
\end{equation}%
The function $\Pi ^{\mathrm{Dim4}}(M^{2},\alpha ,\beta ,\gamma )$ is given
by the expression
\begin{eqnarray}
&&\Pi ^{\mathrm{Dim4}}(M^{2},\alpha ,\beta ,\gamma )=\frac{\langle \alpha
_{s}G^{2}/\pi \rangle C}{384A^{5}\pi ^{4}}\exp \left[ -\frac{A\left(
m_{b}^{2}(\alpha +\beta )-L_{2}m_{c}^{2}\right) }{M^{2}L_{1}\alpha \beta
\gamma }\right] [m_{b}^{2}(\alpha +\beta )-L_{2}m_{c}^{2}]  \notag \\
&&\times \{m_{b}^{4}L_{1}\alpha \beta \gamma (\alpha +\beta )^{2}(\alpha
^{2}+\beta ^{2})+m_{c}^{4}L_{2}^{2}\alpha \beta \gamma (L_{2}+\gamma
)(L_{2}^{2}+2L_{2}\gamma +2\gamma ^{2})  \notag \\
&&-m_{b}^{2}m_{c}^{2}[L_{2}^{4}\alpha ^{2}\beta ^{2}+2L_{2}^{2}\alpha \beta
\gamma (\alpha +\beta +2\alpha ^{3}+5\alpha \beta (\beta -1)+2\beta
^{2}(\beta -1)+\alpha ^{2}(5\beta -2)]  \notag \\
&&+(\alpha +\beta )\gamma ^{4}[\alpha ^{3}+5\alpha ^{2}\beta +\beta
^{3}+\alpha \beta (5\beta -4)]+L_{2}\gamma ^{2}(\alpha ^{5}+\beta
^{4}(-1+\beta +2\gamma )  \notag \\
&&+\alpha ^{4}(-1+11\beta +2\gamma )+2\alpha ^{3}\gamma (-7+13\beta +6\gamma
)+2\alpha ^{2}\beta (3-4\gamma +\beta (-14+13\beta +10\gamma ))  \notag \\
&&+\alpha \beta ^{2}(6-8\gamma +\beta (-14+11\beta +12)\gamma ))))\}.
\end{eqnarray}%
An explicit formula for $\rho ^{\mathrm{Dim4}}(s,\alpha ,\beta ,\gamma )$ is
rather cumbersome, therefore we do not provide it here.

For the vector tetraquark,  the function $\widetilde{\rho }^{\mathrm{pert.}%
}(s,\alpha ,\beta ,\gamma )$ has the following form%
\begin{eqnarray}
&&\widetilde{\rho }^{\mathrm{pert.}}(s,\alpha ,\beta ,\gamma )=\frac{%
N^{2}\Theta ({N)}}{256C^{4}A^{4}\pi ^{6}}\left\{ 4CA^{2}N\left[
5L_{1}^{2}s\alpha ^{2}\beta ^{2}\gamma ^{2}+A^{2}(m_{b}^{2}\alpha \beta
+2m_{c}^{2}L_{1}\gamma )\right] -3L_{1}A^{4}N^{2}\alpha \beta \gamma \right.
\notag \\
&&\left. -12C^{2}(A^{2}m_{c}^{2}+L_{1}s\alpha ^{2}\beta ^{2}\gamma
)(A^{2}m_{b}^{2}+L_{1}^{2}s\alpha \beta \gamma ^{2})\right\} .
\end{eqnarray}%
The function $\widetilde{\Pi }^{\mathrm{Dim4}}(M^{2},\alpha ,\beta ,\gamma )$
is determined by the formula
\begin{eqnarray}
&&\widetilde{\Pi }^{\mathrm{Dim4}}(M^{2},\alpha ,\beta ,\gamma )=\frac{%
\langle \alpha _{s}G^{2}/\pi \rangle C}{768A^{4}\pi ^{4}\alpha \beta \gamma }%
\exp \left[ -\frac{A\left( m_{b}^{2}(\alpha +\beta )-L_{2}m_{c}^{2}\right) }{%
M^{2}L_{1}\alpha \beta \gamma }\right] \left( m_{b}^{2}(\alpha +\beta
)-L_{2}m_{c}^{2}\right) ^{2}  \notag \\
&&\times \{4m_{b}^{4}L_{1}\alpha \beta \gamma (\alpha +\beta )^{2}(\alpha
^{2}+\beta ^{2})+4m_{c}^{4}L_{1}L_{2}^{2}\alpha \beta \gamma (L_{2}+2\gamma
)^{2}-m_{b}^{2}m_{c}^{2}[4L_{2}^{4}\alpha ^{2}\beta ^{2}+L_{2}^{2}\alpha
\beta \gamma  \notag \\
&&\times (11\alpha ^{3}+8\beta (\beta -1)^{2}+8\alpha (\beta -1)(5\beta
-1)+\alpha ^{2}(43\beta -16)]-(\alpha +\beta )\gamma ^{4}(\alpha
^{3}-31\alpha ^{2}\beta  \notag \\
&&+4\alpha \beta (8-7\beta )+4\beta ^{3})+L_{2}\gamma ^{2}(-\alpha
^{5}+\alpha ^{4}(1+40\beta -2\gamma )-4\beta ^{4}(\beta +2\gamma -1)+6\alpha
^{3}\beta (-13  \notag \\
&&+22\beta +10\gamma )+4\alpha \beta ^{2}[10-16\gamma +\beta (-18+7\beta
+12\gamma )]+\alpha ^{2}\beta (40-64\gamma +\beta (-171+123\beta  \notag \\
&&+118\gamma ))))\}.
\end{eqnarray}%
We again refrain from writing down dimension-$4$ component $\widetilde{\rho }%
^{\mathrm{Dim4}}(s,\alpha ,\beta ,\gamma )$ of the spectral density.

\end{widetext}

\end{document}